\documentclass[journal]{IEEEtran}
\ifCLASSINFOpdf
\usepackage[pdftex]{graphicx}
\else
\fi
\usepackage{breakcites}
\usepackage{notoccite}
\usepackage{multirow}
\usepackage{fixltx2e}
\usepackage{url} 
\usepackage{float}
\usepackage{wrapfig} 
\usepackage{graphicx}
\usepackage[noadjust]{cite}
\usepackage[cmex10]{amsmath}
\usepackage{enumitem}
\usepackage{epstopdf}
\usepackage{comment}
\usepackage{etoolbox}
\usepackage{amsfonts}
\usepackage{amsmath}
\usepackage{footnote}
\usepackage{amsthm}
\usepackage{changes}
\usepackage{color}
\usepackage{lipsum}
\definechangesauthor[name={Per cusse}, color=orange]{per}
\setremarkmarkup{(#2)}

\begin{document}
\title{A Survey of Indoor Localization Systems and Technologies}

\author{Faheem~Zafari,~\IEEEmembership{Student Member,~IEEE,}
	 Athanasios~Gkelias,~\IEEEmembership{Senior~Member,~IEEE,}\\
        Kin~K.~Leung,~\IEEEmembership{Fellow,~IEEE}

\thanks{Faheem Zafari,  Athanasios Gkelias and Kin K. Leung are with the Department of Electrical and Electronics Engineering, Imperial College, London, UK e-mail: \{faheem16, a.gkelias, kin.leung\}@imperial.ac.uk}}

\maketitle

\begin{abstract}
	Indoor localization has recently witnessed an increase in interest, due to the potential wide range of services it can provide by leveraging \textit{Internet of Things} (IoT), and ubiquitous connectivity. Different techniques,  wireless technologies and mechanisms have been proposed in the literature to provide  indoor localization services in order to improve the services provided to the users. However, there is a lack of an up-to-date survey paper that incorporates some of the recently proposed accurate and reliable localization systems. In this paper, we aim to provide a detailed survey of different indoor localization techniques such as Angle of Arrival (AoA), Time of Flight (ToF), Return Time of Flight (RTOF), Received Signal Strength (RSS); based on technologies such as WiFi, Radio Frequency Identification Device (RFID), Ultra Wideband (UWB), Bluetooth  and systems that have been proposed in the literature. The paper primarily discusses localization and positioning of human users and their devices. We highlight the strengths of the existing systems proposed in the literature. In contrast with the existing surveys, we also evaluate different systems from the perspective of energy efficiency, availability, cost, reception range, latency, scalability and tracking accuracy. Rather than comparing the technologies or techniques, we compare the localization systems and summarize their working principle.
 We also discuss remaining challenges to accurate indoor localization.   
\end{abstract}

\begin{IEEEkeywords}
Indoor Localization,  Location Based Services,  Internet of Things.
\end{IEEEkeywords}

\IEEEpeerreviewmaketitle

\section{Introduction}
\label{sec:intro}

The wide-scale proliferation of smart phones and other wireless devices in the last couple of years has resulted in a wide range of services including \textit{indoor localization}. 
Indoor localization is the process of obtaining a device or user location in an indoor setting or environment. Indoor device localization has been extensively investigated over the last few decades, mainly in industrial settings and for wireless sensor networks and robotics. However, it is only less than a decade ago since the wide-scale proliferation of smart phones and wearable devices with wireless communication capabilities have made the localization and tracking of such devices synonym to the localization and tracking of the corresponding users and enabled a wide range of related applications and services. User and device localization have wide-scale applications in health sector, industry, disaster management \cite{asimakopoulou2011buildings,borrion2012countering,zelenkauskaite2012interconnectedness}, building management, surveillance and a number of various other sectors. It can also benefit many novel systems such as Internet of Things (IoT) \cite{atzori2010internet}, smart architectures (such as smart cities \cite{su2011smart}, smart buildings \cite{snoonian2003smart}, smart grids \cite{siano2014demand}) and Machine Type Communication (MTC) \cite{taleb2012machine}. 
\par IoT is an amalgamation of numerous heterogeneous technologies and communication standards that intend to provide end-to-end connectivity to billions of devices. Although currently the research and commercial spotlight is on emerging technologies related to the long-range machine-to-machine communications, existing short- and medium-range technologies, such as Bluetooth, Zigbee, WiFi, UWB, etc., will remain inextricable part{s} of the IoT network umbrella. While long-range IoT technologies aim to provide high coverage and low power communication solution, they are incapable to support the high data rate required by various applications in local level. This is why a great number of IoT devices (depending on the underlying application) will utilize more than one communication interface, one for short and one for long range communication.
\par  On the other hand, although long-range IoT technologies have not been designed with indoor localization provision, many IoT applications will require seamless and ubiquitous indoor/outdoor localization and/or navigation of both static and mobile devices. Traditional short-range communication technologies can estimate quite accurately the relative indoor location of an IoT device with respect to some reference points, but the global location (i.e., longitude-latitude geographic coordinates) of these devices remains unknown, unless the global location of the reference points is also known. Emerging long-range IoT technologies can provide an estimate of the global location of a device (since the exact locations of their access points are normally known), however their accuracy deemed very low, especially for indoor environments. We believe that the close collaboration between short- and long-range IoT technologies will be needed in order to satisfy the diverse localization requirements of the future IoT networks and services. This is why in this survey paper we are addressing how both traditional short-range and emerging long-range IoT technologies can be used for localization (even though the latter cannot be explicitly used for indoor localization).
\par Before we start the description of the different localization techniques, technologies and systems, we would like to summarize the various notations and symbols which will be used in this paper in Table \ref{tab:symbols}. Moreover we introduce the following definitions: 
\begin{itemize}
	\item Device based localization (DBL): The process in which the user device uses some \textit{Reference Nodes} (RN) or anchor nodes to obtain its relative location. DBL is primarily used for navigation, where the user needs assistance in navigating around any space. 
	\item Monitor based localization (MBL): The process in which a set of anchor nodes or RNs passively obtains the position of the user or entity connected to the reference node. MBL is primarily used for tracking the user and then accordingly providing different services. 
	\item Proximity Detection: The process of estimating the distance between a user and a \textit{Point of Interest} (PoI). Proximity detection has recently been seen as a reliable and cost effective solution for context aware services \footnote{\textit{Services provided to the user based not only on location, but also the user relevant information such as age, gender, preference etc.}}.  
\end{itemize}
\begin{table*}[]
	\centering
	\scriptsize
	\caption{Notations used throughout the paper}
	\begin{tabular}{|l|p{3.5cm}||p{0.7cm}|p{3.5cm}|}
		\hline
			
{{AoA}} & {{Angle of Arrival}} &  {{BLE}}&  {{Bluetooth Low Energy}}  \\ \hline
		 {{CSI}}& {{Channel State Information}} & {{CFR}} & {{Channel Frequency Response}}  \\ \hline
		{{CSS}} & {{Chirp Spread Spectrum}} & {{DBL}} & {{Device based Localization}}  \\ \hline
	{ {dBm}} & {{decibel-milliwatts}} &{{EKF}} & {{Extended Kalman Filter}}  \\ \hline
		GW   & Gateways  & {{GPS}} & {{Global Positioning System}}  \\ \hline
		{{IoT}} & {{Internet of Things}} & {{ISM}} & {{Industrial, Scientific and Medical}}  \\ \hline
	{{ID}} & {{Identity}} & {{KF}} & {{Kalman Filter}} \\ \hline
	{ {kNN}} &  {{k-Nearest Neighbor}} & {{LoRA}} &  {{Long Range Radio}} \\ \hline
	{{LBS}} &{{Location Based Services}} & {{LoS}} &  {{Line-of-Sight}} \\ \hline
	{{LEDs}} & {{Light Emitting Diodes}} & {{LPWAN}} &  {{Low Power Wide Area Network}} \\ \hline
	MTC	  & Machine-Type Communication  &  {{MAC}} & {{Medium Access Control}} \\ \hline
	 {{MBL}} &  {{Monitor/Mobile Based Localization}} &  {{ns}} &  {{nano second}} \\ \hline
{{PHY}} &  {{Physical Layer}} & {{PF}} &  {{Particle Filter}} \\ \hline
{{PBS}} &  {{Proximity based Systems}} &  {{PoA}} &   {{Phase-of-Arrival}} \\ \hline
 {{RN}} &  {{Reference Node}} &  {{RSSI}} & {{Received Signal Strength Indicator}} \\ \hline
 {{RToF}} &  {{Return Time of Flight}} & RF   & Radio Frequency  \\ \hline
 {{Rx}} &  {{Receiver}} & {{RFID}} &  {{Radio Frequency Identification}} \\ \hline
{{S}} & {{Distance}} &  {{SAR}} &   {{Synthetic Aperture Radar}} \\ \hline 
	{{SVM}} &  {{Support Vector Machine}} &  {{ToF}} &  {{Time of Flight}} \\ \hline
 {{Tx}} & {{Transmitters}} & {{T}} & {{Time}} \\ \hline
{{UHF}} &  {{Ultra-high Frequency}} & UNB   &Ultra-Narrow Band \\ \hline
	{{UWB}} &  {{Ultra-wideband}} &  {{V}} &  {{Propagation Speed}} \\ \hline
 {{VLC}} &  {{Visible Light Communication}} &  {{2D}} &  {{2-Dimensional}} \\ \hline
	 {{3D}} &  {{3-Dimensional}} & {} &   {} \\ \hline
	\end{tabular}
	\label{tab:symbols}
\end{table*}
It is important to differentiate between device and monitor based localization since each of them has different requirements in terms of energy efficiency, scalability and performance. It is also worth mentioning that proximity is another type of localization which requires the relative distance between two objects (or users) of interest instead of their exact location. 
While the first generation of \textit{Location based Services} (LBS)  did not garner significant attention due to its network-centric approach, the second generation of LBS is user-centric and is attracting the interest of researchers around the world \cite{zafari2016microlocation}. Both service providers and end users can benefit from LBS and \textit{Proximity based Services} (PBS). For example, in any shopping mall, the users can use navigation and tracking services to explore the store and get to their desired location. The user can be rewarded by the shop or the mall through discount coupons or promotions based on their location, which will improve the customer experience. The service provider can also benefit from such a system as the  \textit{anonymized} user location data can provide useful insights about the shopping patterns, which can be used to increase their sales. 

\subsection{Existing Indoor Localization Survey Papers}
While the literature contains a number of survey articles \cite{al2011survey,amundson2009survey,gu2009survey,harle2013survey,koyuncu2010survey,liu2007survey,maghdid2016seamless,xiao2016survey,davidson2016survey,ferreira2017localization} on indoor localization, there is a need for an up-to-date survey paper that discusses some of the latest systems and developments \cite{vasisht2016decimeter,kumar2014accurate,xiong2013arraytrack,kotaru2015spotfi,wu2012fila,lim2005zero,goswami2011wigem,sen2012you,nandakumar2012centaur,xiong2015tonetrack,xiong2015pushing} in the field of indoor localization with emphasis on tracking users and user devices. 
Al Nuaimi et al. \cite{al2011survey} provide a discussion on  different indoor localization systems proposed in the literature and highlight challenges such as accuracy that localization systems face. Liu et al. \cite{liu2007survey} provide a detailed survey of various indoor positioning techniques and systems. The paper provides detailed discussion on the technologies and techniques for indoor localization  as well as present some localization systems.  Amundson et al. \cite{amundson2009survey} presents a survey on different localization methods for wireless sensors. The survey primarily deals with WSNs and is for both indoor and outdoor environment. {Davidson et al. \cite{davidson2016survey} provide a survey of different indoor localization methods using smartphone. The primary emphasis of the paper is fingerprinting (radio/magnetic) and smartphone based localization systems. Ferreira et al. \cite{ferreira2017localization} present a detailed survey of indoor localization system for emergency responders. While different localization techniques, and technologies have been discussed, the survey primarily discusses systems that are for emergency response systems.  }

\par  However, the existing surveys do not provide an exhaustive and detailed discussion on the access technologies and techniques that can be used for localization {and only use them for comparing different solutions proposed in literature. Furthermore, most of the existing surveys are specific to a certain domain (such as \cite{davidson2016survey} deals with smartphone based localization, and \cite{ferreira2017localization} deals with emergency responding situations). Therefore, there is a need for a generic survey, which presents a discussion on some of the novel systems that provide high localization accuracy. { For sake of completeness and to make our survey applicable also to readers who have no prior expertise in indoor localization,  we provide a tutorial on basics of techniques and technologies that are used for indoor localization.  }}  {Furthermore, the wide-scale connectivity offered by IoT can also open a wide range of opportunities for indoor localization. It is important to understand  opportunities, and challenges of leveraging the IoT infrastructure for indoor localization.  } In this paper, we present a thorough and detailed survey of different localization  techniques, technologies and systems. We aim to provide the reader with some of the latest localization  systems and also evaluate them from cost, energy efficiency, reception range, availability, latency, scalability, and localization accuracy perspective. {We also attempt to establish a bidirectional link between IoT and indoor localization. } {Our goal is to provide readers, interested in indoor localization, with a comprehensive and detailed insight into different aspects of indoor localization so that the paper can serve as a starting point for further research. }

\subsection{Key Contributions}
\begin{enumerate}
	\item This work provides a detailed survey of different indoor localization systems particularly for user device tracking that has been proposed in the literature between 1997 and 2018. We evaluate these systems using an evaluation framework to highlight their pros and cons. We primarily focus on some of the more recent solutions (from 2013 to 2018)  {some of which have achieved sub-meter accuracy.}
	\item {This work provides a detailed discussion on different technologies that can be used for indoor localization services. We provide the pros and cons of different technologies and highlight their suitability and challenges for indoor localization. }
	\item {We provide an exhaustive discussion on different techniques that can be used with a number of technologies for indoor localization. The discussed techniques rely on the signals emitted by the access technology {(both monitor based and device based localization)} to obtain an estimate of the user location.} 
	\item {Due to the recent increase of interest in  Internet of Things (IoT),  } {we provide a primer on IoT and highlight  indoor localization induced challenges for IoT. We also discuss some of the emerging IoT technologies that are optimized for connecting billions of IoT devices,} {  and analyze their viability for indoor localization. We conclude, on the basis of our analysis, that the emerging IoT technologies are currently not suitable for sub-meter accuracy.  } 
	\item We discuss an evaluation framework that can be used to assess different localization systems. While indoor localization systems are highly application dependent, a generic evaluation framework can help in thoroughly analyzing the localization system. 
	\item This work also discusses some of the existing and potential applications of indoor localization. Different challenges that indoor localization currently faces are also discussed. 

\end{enumerate}
\subsection{Structure of the Paper }
The paper is further structured as follows. 
\begin{itemize}
	\item Section \ref{sec:techniques}: We discuss different techniques such as  RSSI, CSI,  AoA, ToF, TDoA,  RToF,  and PoA for localization in Section \ref{sec:techniques}. We also discuss fingerprinting/scene analysis as it is one of the widely used methods with RSSI based localization. Furthermore, we discussed techniques such as probabilistic methods,  {\emph{Neural Networks}} (NN),  {\emph{k-Nearest Neighbors}} (kNN) and {\emph{Support Vector Machine}} (SVM) that are used with RSSI fingerprints to obtain user location. 
	\item Section \ref{sec:technologies}: We provide different technologies with particular emphasis on wireless technologies that can be used for indoor localization. We primarily discuss WiFi, Bluetooth, Zigbee, RFID, UWB, Visible Light, Acoustic Signals, and ultrasound. The discussion is primarily from localization perspective and we discuss the advantages and challenges of all the discussed technologies. 
		\item {Section \ref{sec:iot}: We present a primer on Internet of Things (IoT). We  list some of the challenges that will arise for IoT due to indoor localization. We  also provide an insight into emerging IoT technologies such as \emph{Sigfox, LoRA, IEEE 802.11ah,} and \emph{weightless} that can be potentially used for indoor localization.  }
	\item Section \ref{sec:framework}:  We present some of the metrics that can be used to evaluate the performance of any localization system. Our evaluation framework consists of metrics such as \emph{availability, cost, energy efficiency, reception range, tracking accuracy,  latency and scalability.}
	\item Section \ref{sec:existing}: We survey various localization systems that have been proposed in literature. We focus on different solutions that have been proposed between 1997 and 2016. Different solutions are evaluated using our evaluation framework. 
	\item Section \ref{sec:applications}: We discuss different possible applications of localization. We highlight the use of localization in  contextual aware location based marketing, health services, disaster management and recovery, security, asset management/tracking and Internet of Things.   
	\item Section \ref{sec:challenges}: We provide a discussion on different challenges that indoor localization systems currently face. We primarily discuss the \emph{multipath effects and noise, radio environment, energy efficiency, privacy and security, cost, negative impact of the localization{\footnote{{negative impact on the basic purpose of the used technology i.e. providing connectivity to the users. As seen in \cite{vasisht2016decimeter}, the throughput of the Wi-Fi AP reduces with increase in the number of users that are to be localized using the AP.}}} on the used technology and the challenges arising due to handovers.  }
	\item Section \ref{sec:conclusions}: We provide the conclusion of the survey. 

\end{itemize}

\section{Localization Techniques}
\label{sec:techniques}
In this section,  {various signal metrics which are widely used for localization will be discussed.}
\subsection{Received Signal Strength Indicator (RSSI)}
The received signal strength (RSS) based approach is one of the simplest and widely used approaches for indoor localization \cite{yang2013rssi,castro2001probabilistic,haeberlen2004practical,krishnan2004system,ladd2005robotics}. The RSS is the actual signal power strength received at the receiver, usually measured in decibel-milliwatts (dBm) or milliWatts (mW). The RSS can be used to estimate the distance between a transmitter (Tx) and a receiver (Rx) device; the higher the RSS value the smaller the distance between Tx and Rx. The absolute distance can be estimated using a number of different signal propagation models given that the transmission power or the power at a reference point is known. RSSI (which is often confused with RSS) is the RSS indicator, a relative measurement of the RSS that has arbitrary units and is mostly defined by each chip vendor. For instance, the Atheros WiFi chipset uses RSSI values between 0 and 60, while Cisco uses a range between 0 and 100. Using the RSSI and a simple path-loss propagation model~\cite{kumar2009distance}, the distance $d$ between Tx and Rx can be estimated from \eqref{eq:pathloss} as
\begin{equation}
\label{eq:pathloss}
\centering 
RSSI=-10n\log_{10}(d)+A,
\end{equation}
where $n$ is the path loss exponent (which varies from 2 in free space to 4 in indoor environments) and  $A$ is the RSSI value at a reference distance from the receiver.

\begin{figure}
	\centering
	\includegraphics[width=0.4\textwidth]{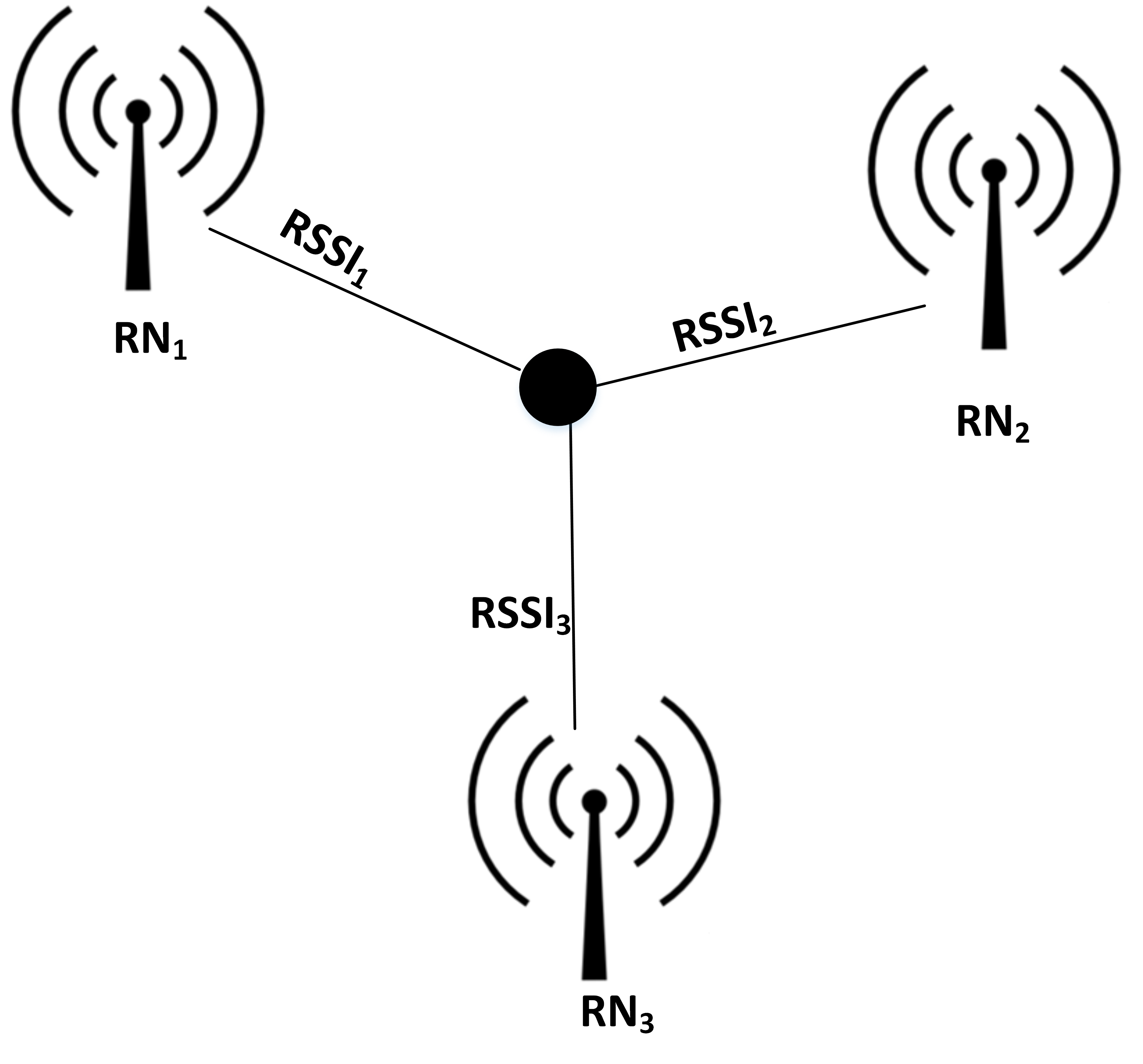}
	\caption{RSSI based localization}
	\protect\label{fig:rssi}
\end{figure}

RSS based localization, in the DBL case, requires trilateration or $N$-point lateration, i.e., the RSS at the device is used to estimate the absolute distance between the user device and at least three reference points; then basic geometry/trigonometry is applied for the user device to obtain its location relative to the reference points as shown in Figure \ref{fig:rssi}. In a similar manner, in the MBL case, the RSS at the reference points is used to obtain the position of the user device. In the latter case, a central controller or ad-hoc communication between anchor points is needed for the total RSS collection and processing. On the other hand, RSS based proximity based services (such as sending marketing alerts to a user when in the vicinity of a retail store), require a single reference node to create a \textit{geofence} \footnote{A virtual fence around any Point of Interest} and estimate the proximity of the user to the anchor node using the path loss estimated distance from the reference point. 
\par {While the RSS based approach is simple and cost efficient, it suffers from poor localization accuracy (especially in non-line-of-sight conditions) due to additional signal attenuation resulting from transmission through walls and other big obstacles and severe RSS fluctuation due to \textit{multipath fading} and indoor noise \cite{yang2013rssi,xiao2013pilot}. Different filters or averaging mechanisms can be used to mitigate these effects. However,  it is unlikely to obtain high localization accuracy without the use of complex algorithms.}

\subsection{Channel State Information (CSI)}

In many wireless systems, such as IEEE 802.11 and UWB, the coherence bandwidth of the wireless channel is smaller than the bandwidth of the signal which makes the channel frequency selective (i.e., different frequencies exhibit different amplitude and phase behavior). Moreover, in multiple antennae transceivers, the channel frequency responses for each antennae pairs may significantly vary (depending on the antennae distance and signal wavelength). While RSS has been widely used due to its simplicity and low hardware requirements, it merely  provides an estimate of the average amplitude over the whole signal bandwidth and the accumulated signal over all antennae. These make RSS susceptible to multipath effects and interference and causes high variability over time. 

On the other hand, the Channel Impulse Response (CIR) or its Fourier pair, i.e., the Channel Frequency Response (CFR), which is normally delivered to upper layers as channel state information (CSI), has higher granularity than the RSS as it can capture both the amplitude and phase responses of the channel in different frequencies and between separate transmitter-receiver antennae pairs {\cite{yang2013rssi}}. In general, the CSI is a complex  {{quantity}} and can be written in a polar form as
\begin{equation}
H(f)=|H(f)|e^{j\angle H(f)},
\end{equation}
where, $|H(f_i)|$ is the amplitude (or magnitude) response and $\angle H(f_i)$ is the phase response of the frequency $f_i$ of the channel. Nowadays, many IEEE 802.11 NICs cards can provide subcarrier-level channel measurements for Orthogonal Frequency Division Multiplexing (OFDM) systems which can be translated into richer multipath information, more stable measurements and higher localization accuracy.

\subsection{{Fingerprinting/Scene Analysis}}
{Scene analysis based localization techniques usually require environmental survey to obtain fingerprints or features of the environment  where the localization system is to be used \cite{zafari2016microlocation,youssef2005horus}. Initially, different RSSI measurements are collected during an offline phase. Once the system is deployed, the online measurements (obtained during real-time) are compared with the offline measurements to estimate the user location. Usually the fingerprints or features are collected in form of RSSI or CSI. There are a number of algorithms available that can be used to match the offline measurements with online measurement, some of which are discussed below. }
\paragraph{{Probabilistic methods}}
{Probabilistic methods rely on the likelihood of the user being in position \textit{`x'} provided the RSSI values, obtained in online phase, are \textit{`y'}. Suppose that the set of location candidates  L is $\emph{L}=\{L_1, L_2, L_3,...., L_m\}$. For any observed online RSSI value vector $O$, user/device location will be $L_j$ if}
\begin{equation}
{P(L_j|O)>P(L_k|O) \;\;for\;\; j,k=1,2,3,.....,m\;\; k\neq j}
\label{eq:probmethod}
\end{equation}
{Equation \eqref{eq:probmethod} shows that a user  {will} be classified in location $L_j$ if its likelihood is higher than any other location. If $P(L_j)=P(L_k)$ for $j,k=1,2,3,.....m$, then using Bayes' theorem, we can obtain the likelihood probability of the observation signal vector being $O$ given that the user is in location $L_j$ as $P(O|L_j)$. Mathematically, the user would be classified in the location $L_j$ if}
\begin{equation}
{P(O|L_j)>P(O|L_k) \;\;for\;\; j,k=1,2,3,.....,m\;\; k\neq j}
\label{eq:probmethod2}
\end{equation}
{The likelihood can be calculated using histogram, and kernel approaches \cite{liu2007survey}.   For independent RNs in space, the  likelihood of user location can be calculated using the product of the likelihoods of all RNs.} 
\par {As described above, fingerprinting methods are using the online RSSI or CSI measurements to map the user/device position on a discrete grid; each point on this grid corresponds to the position in space where the corresponding offline measurements (i.e., fingerprints) were obtained. Therefore, fingerprinting provides discrete rather than continuous estimation of the user/device location. Theoretically, the location estimation granularity can be increased by reducing the distance between the offline measurement points (i.e., increasing the density of the grid) to the point where almost continuous location estimation is obtained. However, in this case, the difference in the signal strength between two neighbor points will become much smaller than the typical indoor signal variations (due to the channel statistics and measurement noise), which makes the estimation of the correct point almost impossible. Therefore, there is an important tradeoff between the fingerprinting position granularity and the probability of successful location estimation which needs to be taken into consideration when the fingerprinting locations are chosen. It is also worth mentioning that while the fingerprinting and scene analysis techniques can provide accurate localization estimations, since they depend on offline and online measurements at different time instances, they are very susceptible to changes of the environment over time.}
\paragraph{{{Artificial} Neural Networks}}
{{Artificial} Neural networks ({A}NN) are used in a number of classification and forecasting scenarios. For localization, the NN has to be trained using the RSSI values and the corresponding coordinates that are obtained during the offline phase  {\cite{altini2010bluetooth}}. Once the ANN is trained, it can then be used to obtain the user location based on the online RSSI measurements. The Multi-Layer Perceptron (MLP) network with one hidden node layer is one of the commonly used ANN for localization \cite{liu2007survey}. In MLP based localization, 
	an input vector of the RSSI  measurements is multiplied with the input weights and added into an input layer bias, provided that bias is selected. The obtained result is then put into hidden layer's transfer function. The product of  the transfer function output and the trained hidden layer weights is added to the hidden layer bias (if bias is chosen). The obtained output is the estimated user location. }

\paragraph{{k-Nearest Neighbor (kNN)}}
{The k-Nearest Neighbor (kNN) algorithms relies on the online RSSI to obtain the k-nearest matches (on the basis of offline RSSI measurements stored in a database) of the known locations using root mean square error (RMSE) \cite{liu2007survey}. The nearest matches are then averaged to obtain an estimated location of the device/user. A weighted kNN is used if the distances are adopted as weights in the signal space, otherwise a non-weighted kNN is used. }
\paragraph{{Support Vector Machine (SVM)}}
{Support vector machine is an attractive approach for classifying data as well as regression. SVM is primarily used for machine learning (ML) and statistical analysis and has high accuracy. As highlighted in \cite{liu2007survey}, SVM can also be used for localization using offline and online RSSI measurements.}

\subsection{Angle of Arrival (AoA)}
\begin{figure}
	\centering
	\includegraphics[width=0.5\textwidth]{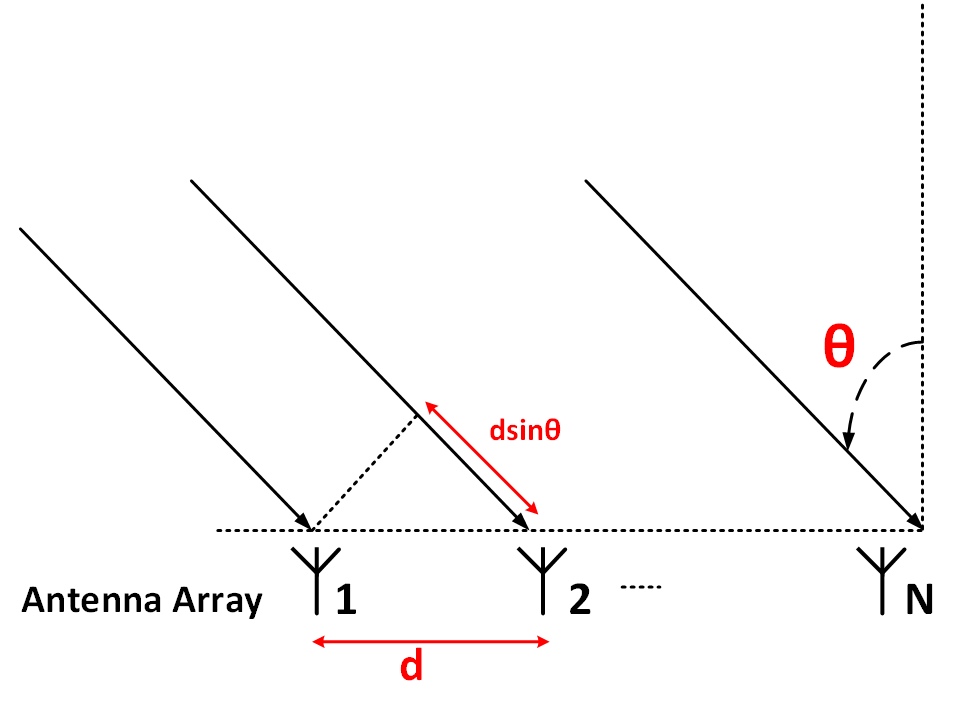}
	\caption{AoA based localization}
	\protect\label{fig:aoa}
\end{figure}

Angle of Arrival (AoA) based approaches use antennae arrays~\cite{xiong2013arraytrack} (at the receiver side) to estimate the angle at which the transmitted signal impinges on the receiver by exploiting and calculating the time difference of arrival at individual elements of the antennae array. The main advantage of AoA is that the device/user location can be estimated with as low as two monitors in a 2D environment, or three monitors in a 3D environment respectively. 
Although AoA can provide accurate estimation when the transmitter-receiver distance is small, it requires more complex hardware and careful calibration compared to RSS techniques, while its accuracy deteriorates with increase in the transmitter-receiver distance where a slight error in the angle of arrival calculation is translated into a huge error in the actual location estimation~\cite{kumar2014accurate}. Moreover, due to multipath effects in indoor environments the AoA in terms of line of sight (LOS) is often hard to obtain. \textit{Figure~\ref{fig:aoa} shows how AoA can be used to estimate the user location {(as the angles at which the signals are received by the antenna array can help locate the user device.)}}.

\subsection{Time of Flight (ToF)}
\begin{figure}
	\centering
	\includegraphics[width=0.5\textwidth]{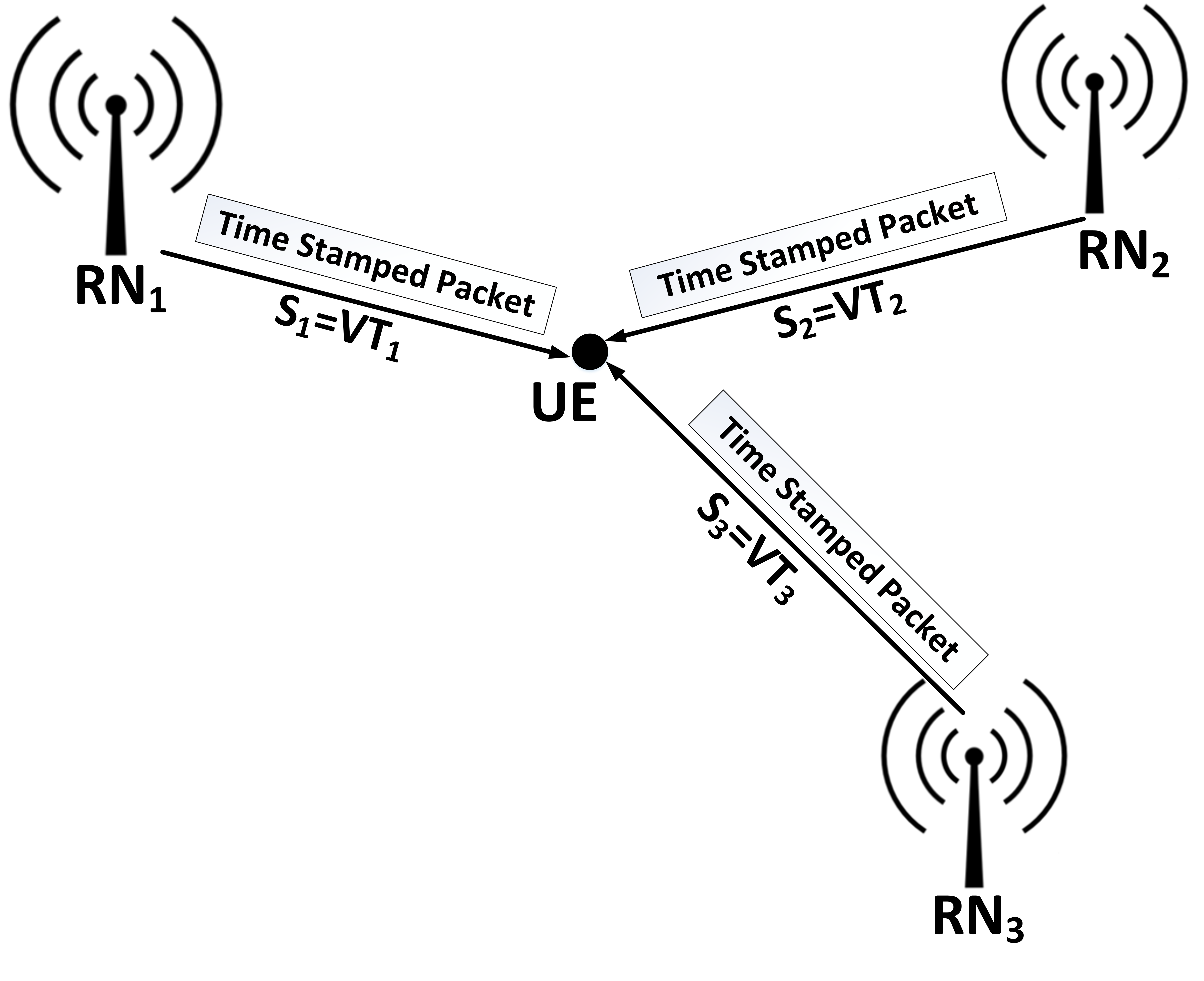}
	\caption{ToF based {user equipment (UE)} localization}
	\protect\label{fig:toa}
\end{figure}

Time of Flight (ToF) or Time of Arrival (ToA) exploits the signal propagation time to calculate the distance between the transmitter Tx and the receiver Rx {\cite{dargie2010fundamentals}}. The ToF value multiplied by the speed of light $c=3\times 10^8$ m/sec provides the physical distance between Tx and Rx. In Figure~\ref{fig:toa}, the ToF from three different reference nodes is used to estimate the distances between the reference nodes and the device. Basic geometry can be used to calculate the location of the device with respect to the access points. Similar to the RSS, the ToF values can be used in both the DBL and MBL scenarios. 

ToF requires strict synchronization between transmitters and receivers and, in many cases, timestamps to be transmitted with the signal (depending on the underlying communication protocol). The key factors that affect ToF estimation accuracy are the signal bandwidth and the sampling rate. Low sampling rate (in time) reduces the ToF resolution since the signal may arrive between the sampled intervals. Frequency domain super-resolution techniques are commonly used to obtain the ToF with high resolution from the channel frequency response. In multipath indoor environments, the larger the bandwidth, the higher the resolution of ToF estimation. Although large bandwidth and super-resolution techniques can improve the performance of ToF, still they cannot eliminate significant localization errors when the direct line of sight path between the transmitter and receiver is not available. {{This is because the obstacles deflect the emitted signals, which then traverse through a longer path causing an increase in the time taken for the signal to propagate from Tx to Rx. }}
{Let $t_1$ be the time when Tx $i$ sends a message to the Rx $j$ that receives it at $t_2$ where $t_2=t_1+t_p$ ($t_p$ is the time taken by the signal to traverse from Tx to Rx) \cite{dargie2010fundamentals}. So the distance between the $i$ and $j$ can be calculated using Equation \eqref{eq:ToF} }
\begin{equation}
\centering 
\label{eq:ToF}
{D_{ij}={(t_2-t_1)}\times{v}}
\end{equation}
{where $v$ is the signal velocity.} 
\subsection{Time Difference of Arrival (TDoA)}

\begin{figure}
	\centering
	\includegraphics[width=0.48\textwidth]{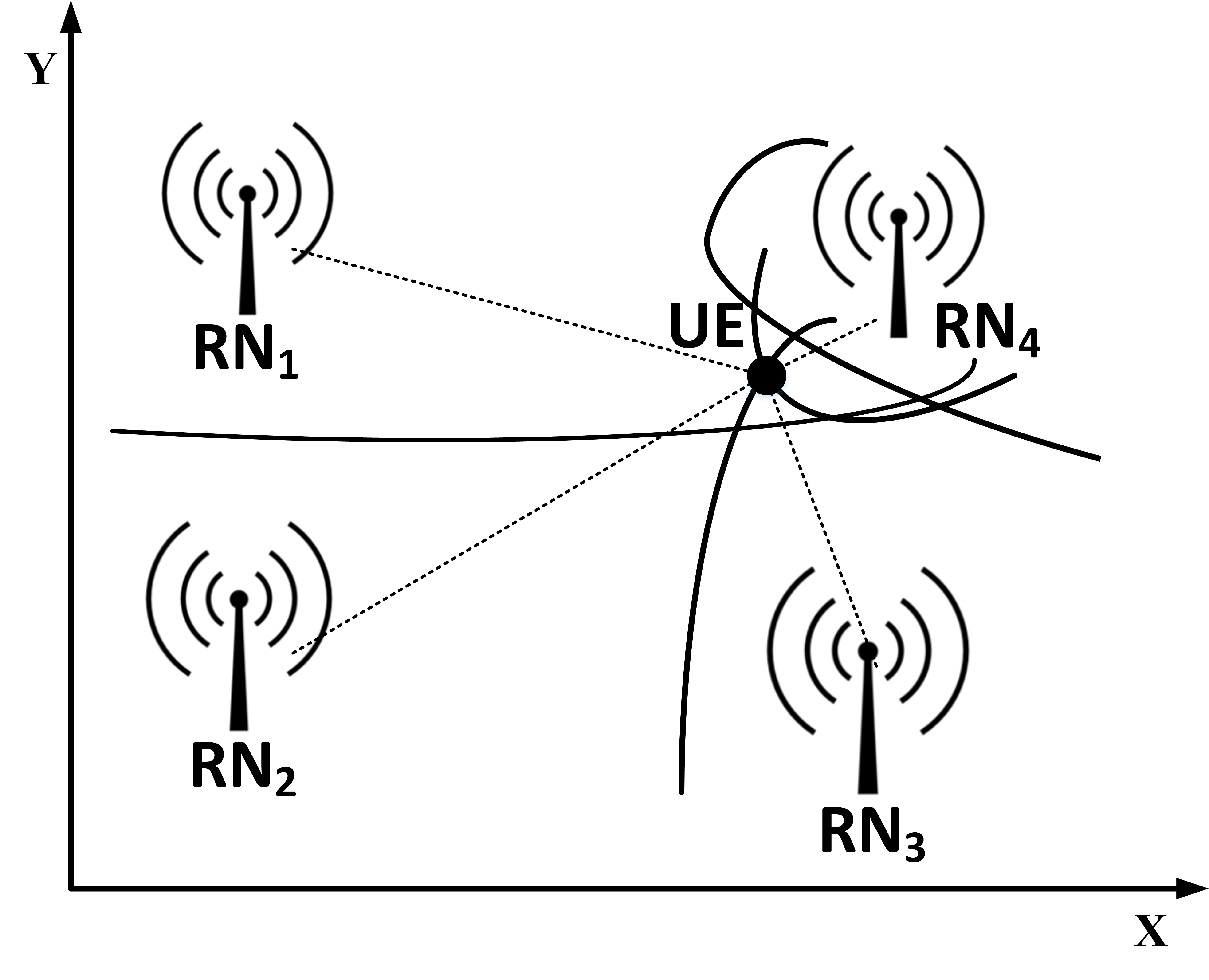}
	\caption{TDoA based localization and proximity detection}
	\protect\label{fig:tdoa}
\end{figure}

Time Difference of Arrival (TDoA) exploits the difference in signals propagation times from different transmitters, measured at the receiver. This is different from the ToF technique, where the absolute signal propagation time is used. The TDoA measurements ($T_{D(i,j)}$ - from transmitters $i$ and $j$) are converted into physical distance values $L_{D(i,j)}=c \cdot T_{D(i,j)}$, where $c$ is the speed of light. The receiver is now located on the hyperboloid given by Eq.\eqref{eq:tdoa} 
\begin{align}\nonumber
\label{eq:tdoa}
L_{D(i,j)}&=\sqrt{(X_i-x)^2+(Y_i-y)^2+(Z_i-z)^2}\\  &
-\sqrt{(X_j-x)^2+(Y_j-y)^2+(Z_j-z)^2},
\end{align}
where $(X_i,Y_i,Z_i)$ are the coordinates of the transmitter/reference node $i$ and $(x,y,z)$ are the coordinates of the receiver/user. The TDoA from at least three transmitters is needed to calculate the exact location of the receiver as the intersection of the three (or more) hyperboloids. The system of hyperbola equations can be solved either through linear regression \cite{liu2007survey} or by linearizing the equation using Taylor-series expansion.  {Figure \ref{fig:tdoa} shows how four different RNs can be used to obtain the 2D location of any target. Figure shows the hyperbolas  formed as a result of the measurements obtained from the RNs to obtain the user location (black dot)}. The TDoA estimation accuracy depends (similar to the ToF techniques) on the signal bandwidth, sampling rate at the receiver and the existence of direct line of sight between the transmitters and the receiver. Strict synchronization is also required, but unlike ToF techniques where synchronization is needed between the transmitter and the receiver, in the TDoA case only synchronization between the transmitters is required. 

\subsection{Return Time of Flight (RToF)}
RToF measures the round-trip (i.e., transmitter-receiver-transmitter) signal propagation time to estimate the distance between Tx and Rx {\cite{dargie2010fundamentals}}. The ranging mechanisms for both ToF and RToF are similar; upon receiving a signal from the transmitter, the receiver responds back to the transmitter, which then calculates the total round-trip ToF. The main benefit of RToF is that a relatively moderate clock synchronization between the Tx and the Rx is required, in comparison to ToF. However, RToF estimation accuracy is affected by the same factors as ToF (i.e., sampling rate and signal bandwidth) which in this case is more severe since the signal is transmitted and received twice. Another significant problem with RToF based systems is the response delay at the receiver which highly depends on the receiver electronics {and protocol overheads}. The latter one can be neglected if the propagation time between the transmitter and receiver is large compared to the response time, however the delay cannot be ignored in short range systems such as those used for indoor localization. 
{Let $t_1$ be the time when Tx $i$ sends a message to the Rx $j$ that receives it at $t_2$ where $t_2=t_1+t_p$. $j$, at time $t_3$, transmits a signal back to $i$ that receives it at $t_4$  So the distance between the $i$ and $j$ can be calculated using Equation \eqref{eq:RToF} {\cite{dargie2010fundamentals}} }
\begin{equation}
\centering 
\label{eq:RToF}
{D_{ij}=\frac{(t_4-t_1)-(t_3-t_2)}{2}\times v}
\end{equation}

\subsection{Phase of Arrival (PoA)}
\begin{figure}
	\centering
	\includegraphics[width=0.45\textwidth]{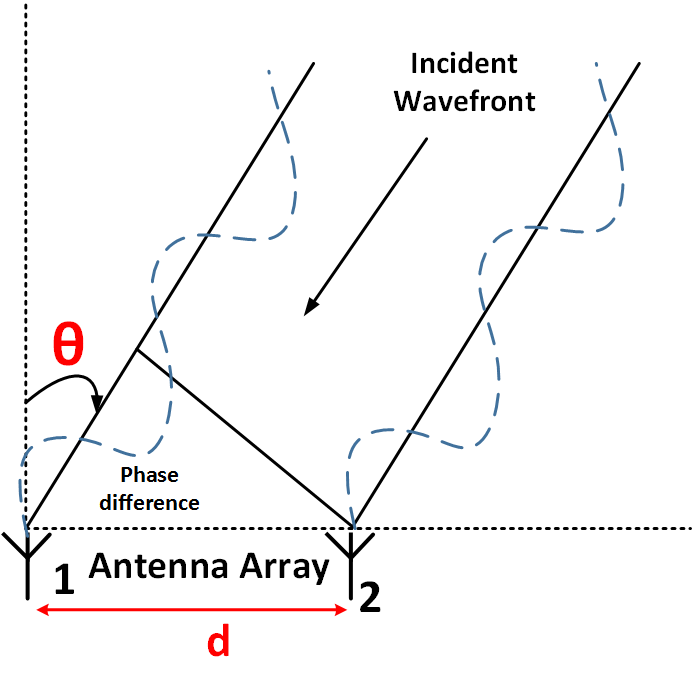}
	\caption{PoA based localization}
	\protect\label{fig:poa}
\end{figure}

PoA based approaches use the phase or phase difference of carrier signal to estimate the distance between the transmitter and the receiver. A common assumption for determining the phase of signal at receiver side is that the signals transmitted from the anchor nodes (in DBL), or user device (in MBL) are of pure sinusoidal form having same frequency and zero phase offset.   
There are a number of techniques available to estimate the range or distance between the Tx and Rx using PoA. One technique is to assume that there exists a finite transit delay $D_i$ between the Tx and Rx, which can be expressed as a fraction of the signal wavelength. {As seen in Figure \ref{fig:poa}, the incident signals arrive  with a phase difference at different antenna in the antenna array, which can be used to obtain the use{r} location.}  A detailed discussion on PoA-based range estimation is beyond the scope of the paper. Therefore interested readers are referred to \cite{povalavc2010phase,scherhaufl2013phase}. 
Following range estimation, algorithms used for ToF can be used to estimate user location. If the phase difference between two signals transmitted from different anchor points is used to estimate the distance,  TDoA based algorithms can be used for localization. PoA can be used in conjunction with RSSI, ToF, TDoA to improve the localization accuracy and enhance the performance of the system.  The problem with PoA based approach is that it requires line-of-sight for high accuracy, which is rarely the case in indoor environments.  

 Table \ref{tab:techniques} provides a summary of the discussed techniques for indoor localization and discusses the advantages and disadvantages of these techniques. { Interested readers are referred to \cite{dargie2010fundamentals} for detailed discussion on these localization techniques. }

\begin{table*}[]
\centering
\caption{Advantages and Disadvantages of different localization techniques}
\begin{tabular}{|l|p{7cm}|p{7cm}|}
\hline
\textbf{Technique} & \textbf{Advantages} & \textbf{Disadvantages} \\ \hline
RSSI               &    Easy to implement, cost efficient, can be used with a number of technologies                  &  Prone to multipath fading and environmental noise, lower localization accuracy, can require  {{fingerprinting}}                     \\ \hline
CSI                &  More robust to multipath and indoor noise,                   & It is not easily available on off-the-shelf NICs                      \\ \hline
AoA	&   Can provide high localization accuracy, does not require any fingerprinting                   &    Might require directional antennas and complex hardware, requires comparatively complex algorithms and performance deteriorates with increase in distance between the transmitter and receiver                    \\ \hline
ToF              &    Provides high localization accuracy, does not require any fingerprinting                 &  Requires time synchronization between the transmitters and receivers, might require time stamps and multiple antennas at the transmitter and receiver. Line of Sight is mandatory for accurate performance.                      \\ \hline
TDoA                &  Does not require any fingerprinting, does not require clock synchronization among the device and RN                   &   Requires clock synchronization among the RNs, might require time stamps, requires larger bandwidth                     \\ \hline
RToF	&   Does not require any fingerprinting, can provide high localization accuracy                  &   Requires clock synchronization, processing delay can affect performance in short ranger measurements                      \\ \hline
PoA               &    Can be used in conjunction wit{{h}} RSS, ToA, TDoA to improve the overall localization accuracy                 & Degraded performance in the absence of line of sight                       \\ \hline
 {Fingerprinting}            &    {Fairly easy to use}            &        {New fingerprints are required even when there is a minor variation in the space}                  \\ \hline
		
	\end{tabular}
		\label{tab:techniques}
\end{table*}

\section{Technologies for localization}
\label{sec:technologies}
In this section, several existing technologies which have been used to provide indoor localization services will be 
presented and discussed. Radio communication technologies, such as, IEEE 802.11, Bluetooth, Zigbee,  RFID and 
Ultra-Wideband (UWB), will be presented first, followed by visible light and acoustic based technologies. Finally, 
several emerging technologies which can be also used as localization enablers will be discussed. 
While there  {{are}} a number of localization systems based on camera/vision technologies, such systems are beyond the scope 
of this survey and will not be discussed here.   
\subsection{WiFi}

The IEEE 802.11 standard, commonly known as WiFi, operates in the \textit{Industrial, Scientific, and Medical} (ISM) band 
and is primarily used to provide networking capabilities and connection to the Internet to different devices in private, public 
and commercial environments. Initially, WiFi had a reception range of about 100 meters \cite{liu2007survey} which has now 
increased to about 1 kilometer (km) \cite{centenaro2015long,adame2014ieee} in IEEE 802.11ah (primarily optimized 
for IoT services). 

Most of the current smart phones, laptops and other portable user devices are WiFi enabled, which makes WiFi an ideal candidate 
for indoor localization and one of the most widely studied localization  {{technologies}} in the literature \cite{paul2009rssi,kumar2014accurate,kotaru2015spotfi,vasisht2016decimeter,xiong2013arraytrack,xiao2013pilot,jiang2012ariel}, \cite{woo2011application,chintalapudi2010indoor,liu2012push,liu2011wifi,feng2012received,cypriani2009open,zou2014online,ciurana2007wlan},\cite{hoang2013parameter,kang2012improved}. 
Since existing WiFi access points can be also used as reference points for signal collection~\cite{kumar2014accurate}, basic localization systems 
(that can achieve reasonable localization accuracy) can be built without the need for additional infrastructure. 

However, existing WiFi networks are normally deployed  for communication (i.e., to maximize data throughput and 
network coverage) rather than localization  purposes and therefore novel and efficient algorithms are required to improve their localization accuracy. Moreover, the uncontrolled interference in the ISM band has been shown to affect the localization accuracy \cite{faheemthesis}. The aforementioned RSS, CSI, ToF and AoA techniques (and any combination of them - i.e., hybrid methods) can be used to provide WiFi based localization services.  Recent WiFi based localization systems \cite{kumar2014accurate,vasisht2016decimeter,kotaru2015spotfi}, details of which are given in Section \ref{sec:existing},  have achieved median localization accuracy as high as 23cm \cite{xiong2013arraytrack}. For detailed information about WiFi, readers are referred to \cite{lee2008broadband}.

\subsection{Bluetooth}
Bluetooth (or IEEE 802.15.1) consists of the physical and MAC layers specifications for connecting different fixed or moving wireless devices within a certain personal space. The latest version of Bluetooth, i.e., Bluetooth Low Energy (BLE), also known as Bluetooth Smart, can provide an improved data rate of 24Mbps and coverage range of 70-100 meters with higher energy efficiency, as compared to older versions~\cite{zafari2016microlocation}. While BLE can be used with different localization techniques such as RSSI, AoA, and ToF, most of the existing BLE based localization solutions rely on RSS based inputs as RSS based sytems are less complex. The reliance on RSS based inputs limits its localiztion accuracy. 
 Even though BLE in its original form can be used for localization (due to its range, low cost and energy consumption), two BLE based protocols, i.e., \textit{iBeacons} (by Apple Inc.) and \textit{Eddystone} (by Google Inc.), have been recently proposed, primarily for context aware proximity based services. 

Apple announced iBeacons in the \textit{World Wide Developer Conference} (WWDC) in 2013 \cite{ibeacondeveloper}. The protocol is specifically designed for proximity detection and proximity based services. The protocol allows a BLE enabled device (also known as iBeacon or beacon) to transmit beacons or signals at periodic interval. The beacon message consists of a mandatory 16 byte Universally Unique Identifier (UUID)\footnote{It is the universal identifier of the beacon. Any organization \textit{`x'} that intends to have an iBeacon based system will have a constant UUID.} and optional 2 byte major\footnote{The organization x can use the major value to differentiate its store in city y from city z.} and minor values\footnote{Any store x in city y can have different minor values for the beacons in different lanes or sections of the store.}. Any BLE enabled device, that has a proprietary application to listen to the beacons picks up the beacon messages and uses RSSI to estimate the proximity between the iBeacon device and the user. Based on the strength of the RSSI, the user is classified in \textit{immediate}~($<$1m), \textit{near}~(1-3m), \textit{far}~($>$3m) and \textit{unknown} regions. 

\begin{figure}
	\centering
	\includegraphics[width=0.4\textwidth]{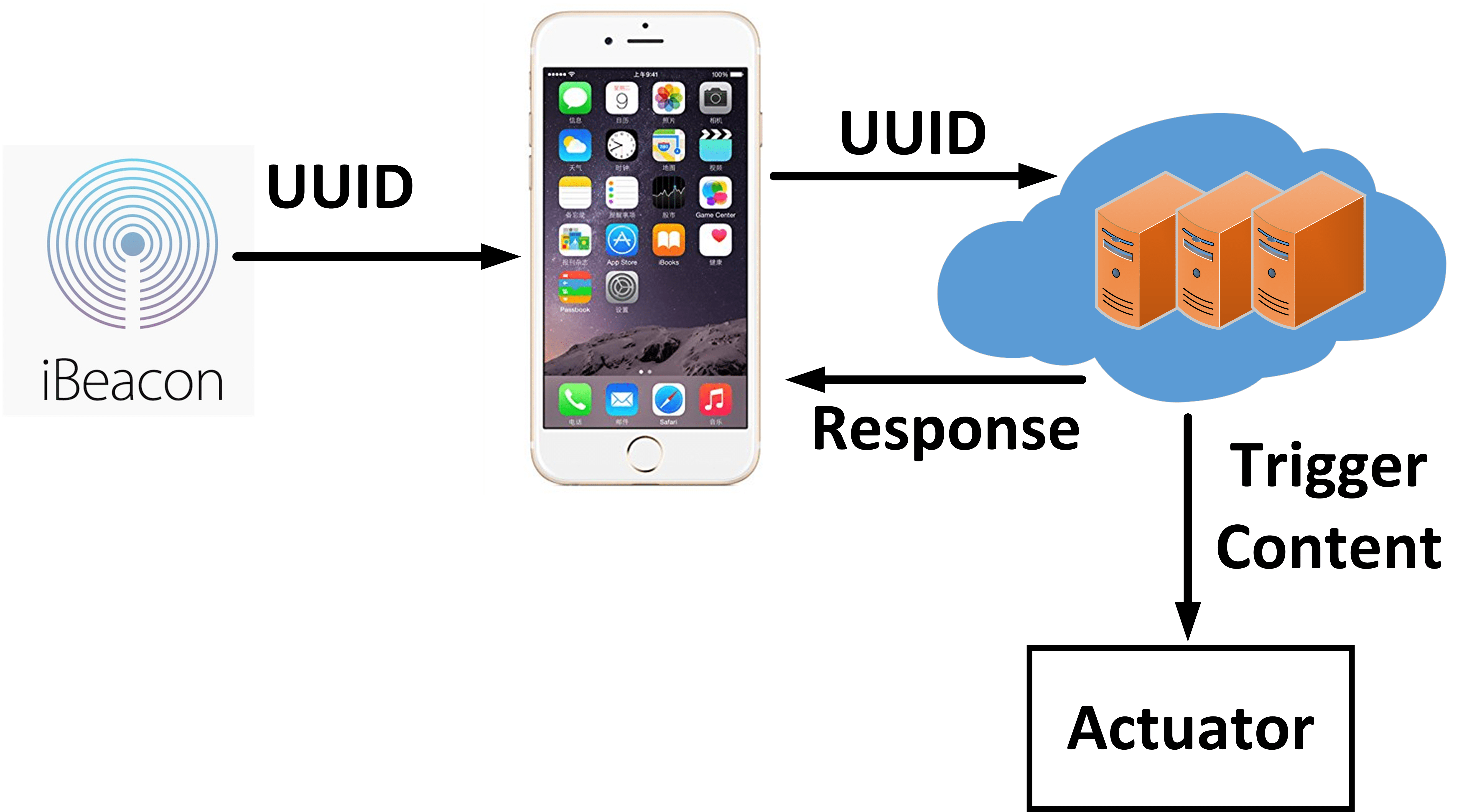}
	\caption{Typical architecture for iBeacon based systems}
	\protect\label{fig:ibeaconarc}
\end{figure}

The schematic of a typical beacon architecture is depicted in Figure \ref{fig:ibeaconarc}. After receiving a message from the iBeacon, the user device consults a server or the cloud to identify the action affiliated with the  received beacon. The action might be to send a discount coupon to be received by the user device, to open a door or to display some interactive content on a monitor (actuator) based on the user's proximity to some beacon or another entity, etc. 
\par
A fundamental constraint of iBeacons (imposed by Apple) is that only the average RSSI value is reported to the user device every one second, even though the beacons are transmitted at 50 ms intervals. This is to account for the variations in the instantaneous RSS values on the user device. However, this RSS averaging and reporting delay can impose significant challenges to real-time localization. While the motive behind iBeacons was to provide proximity detection, it has also been used for indoor localization, details of which can be found in the next section. 
\subsection{Zigbee}
Zigbee is built upon the IEEE 802.15.4 standard that is concerned with the physical and MAC layers for low cost, low data rate and energy efficient personal area networks \cite{baronti2007wireless}.  Zigbee defines the higher levels of the protocol stack and is basically used in wireless sensor networks. The Network Layer in Zigbee is responsible for multihop routing and network organization while the application layer is responsible for distributed communication and development of application. While Zigbee is favorable for localization of sensors in WSN, it is not readily available on majority of the user devices, hence it is not favorable for indoor localization of users.  

\subsection{Radio Frequency Identification Device (RFID)}
RFID is primarily intended for transferring and storing data using electromagnetic transmission from a transmitter to any Radio Frequency (RF) compatible circuit \cite{holm2009hybrid}. An RFID system consists of a reader that can communicate with RFID tags. The RFID tags emit data that the RFID reader can read using a predefined RF and protocol, known to both the reader and tags a  {{priori}}. 
There are two basic types of RFID systems 
\begin{itemize}
	\item \textit{Active RFID}: Active RFIDs operate in the Ultra High Frequency (UHF) and microwave frequency range. They are connected to a local power source, periodically transmit their ID and can operate at hundreds of meters from the RFID reader. Active RFIDs can be used for localization and object tracking as they have a reasonable range, low cost and can be easily embedded in the tracking objects. However, the active RFID technology cannot achieve sub-meter accuracy and it is not readily available on most portable user devices. 
	\item \textit{Passive RFID}: Passive RFIDs are limited in communication range (1-2m)  and can operate without battery. They are smaller, lighter and cost less than the active ones; they can work in the low, high, UHF and microwave frequency range. Although they can be used as an alternative to bar-codes, especially when the tag is not within the line of sight of the reader, their limited range make them unsuitable for indoor localization. They can be used for proximity based services using \textit{brute force} approaches\footnote{Increasing the number of tags deployed in any space}, but this will still require modifications to the existing procedure used by passive RFIDs such as transmitting an ID that can be used to identify the RFID and help 
\end{itemize} 

\subsection{Ultra Wideband (UWB)}
In UWB, ultra short-pulses with time period of $<$1 nanosecond (ns) are transmitted over a large bandwidth ($>$500MHz), in the frequency range from 3.1 to 10.6GHz, using a very low duty cycle \cite{liu2007survey} which results in reduced power consumption. The technology has been primarily used for short-range communication systems, such as PC peripherals, and other indoor applications. UWB has been a particularly attractive technology for indoor localization because it is immune to interference from other signals (due to its drastically different signal type and radio spectrum), while the UWB signal (especially the low frequencies included in the broad range of the UWB spectrum) can penetrate a variety of materials, including walls (although metals and liquids can interfere with UWB signals). Moreover, the very short duration of UWB pulses make them less sensitive to  {{multipath}} effects, allowing the identification of the main path in the presence of multipath signals and providing accurate estimation of the ToF, that has been shown to achieve localization accuracy up to 10cm~\cite{decawave}. 

However, the slow progress in the UWB standard development (although UWB has been initially proposed for use in personal area networks PANs), has limited the use of UWB in consumer products and portable user devices in particular as standard. Since, an in-depth discussion of UWB is beyond the scope of this paper, readers are referred to \cite{gezici2005localization,oppermann2005uwb} for further details. 

\subsection{Visible Light}
\textit{Visible Light Communication} (VLC) is an emerging technology for high-speed data transfer~\cite{kuo2014luxapose} that uses visible light between 400 and 800THz, modulated and emitted primarily by \textit{Light Emitting Diodes} (LEDs). Visible light based localization techniques use light sensors to measure the position and direction of the LED emitters. In other words, the LEDs (acting like the iBeacons) transmit the signal, which when picked up by the receiver/sensor can be used for localization. For visible light, AoA is considered the most accurate localization technique~\cite{armstrong2013visible,kuo2014luxapose}. The advantage of visible light based localization is its wide scale proliferation (perhaps even more than WiFi). However, a fundamental limitation is that line of sight between the LED and the sensor(s) is required for accurate localization. 

\begin{table*}[]
	\centering
	\caption{Summary of different wireless technologies for localization}
	\begin{tabular}{|p{2.4cm}|p{1.7cm}|p{1.7cm}|l|p{4cm}|p{4cm}|}
		\hline
		\textbf{Technology}    & \textbf{\begin{tabular}[c]{@{}l@{}}Maximum \\ Range\end{tabular}} & \textbf{\begin{tabular}[c]{@{}l@{}}Maximum\\ Throughput\end{tabular}} & \textbf{\begin{tabular}[c]{@{}l@{}}Power \\ Consumption\end{tabular}} & \textbf{Advantages}                                                                                                                                     & \textbf{Disadvantages}                                                                                                                   \\ \hline
		\textbf{IEEE 802.11 n $\textsuperscript{ \cite{Intelwirless}}$} & 250 m outdoor                                                     & 600 Mbps                                                              & Moderate                                                              & \multirow{3}{*}{\begin{tabular}[c]{@{}l@{}}Widely available,  high  accuracy,\\ does not 
				require\\ complex extra hardware \end{tabular}} & \multirow{3}{*}{\begin{tabular}[c]{@{}l@{}}Prone to noise, requires\\ complex processing \\ algorithms\end{tabular}} \\
		\textbf{802.11 ac}     & 35 m indoor & 1.3 Gbps  & Moderate     &   &    \\
		\textbf{802.11 ad}     & couple of meters    & 4.6 Mbps   & Moderate    &                                                                                                                                                         &  \\ \hline
		\textbf{UWB $\textsuperscript{ \cite{uwbreport}}$}    &   10-20m         &  460 Mbps  & Moderate                        &      Immune to interference, provides high accuracy,                                                                                                                                                                                             &    Shorter range, requires extra hardware on different user devices, high cost                                                                                                                                       \\ \hline
		\textbf{Acoustics}     &     Couple of meters                                                     &         &           Low-Moderate                                                            &    Can be used for proprietary applications, can provide high accuracy                                                                                                                                                     &  Affected by sound pollution, requires extra anchor points or hardware                                                                                                                                      \\ \hline
		\textbf{RFID $\textsuperscript{ \cite{nadlerpresence}}$}          &     200 m                                                              &   1.67 Gbps         &   Low              & Consumes low power, has wide range                                                                                                                                                        &  Localization accuracy is low                                                                                                                                      \\ \hline
		\textbf{Bluetooth $\textsuperscript{ \cite{ergen2004zigbee}}$}           &  100m                                                                 &   24 Mbps                   &      Low               &                                High throughput, reception range, low energy consumption                                                                                                                         &         Low localization accuracy, prone to noise                                                                                                                                \\ \hline
		\textbf{Ultrasound $\textsuperscript{ \cite{singer2016mbps}}$}    & Couple-tens of meters &  30 Mbps      & Low-moderate                      &  Comparatively less absorption                                                                                                                    &   High dependence on sensor placement                                                                                                                                      \\ \hline
		\textbf{Visible Light $\textsuperscript{ \cite{visiblelight}}$} &  1.4 km  & 10 Gbps $\textsuperscript{\cite{vlcspeed}}$  &    Relatively higher   & Wide-scale availability, potential to provide high accuracy, multipath-free  & Comparatively higher power consumption, range is affected by obstacles, primarily requires LoS                                                                                                                                         \\ \hline
		\textbf{SigFox $\textsuperscript{ \cite{centenaro2015long}}$}        &     50 km      &  100 bps    & Extremely low          &  Wide reception range, low energy consumption                                                                                                                                                       & Long distance between base station  and device, sever outdoor-to-indoor signal attenuation due to building walls \\ \hline
		\textbf{LoRA $\textsuperscript{ \cite{centenaro2015long}}$}          &    15 km                                                             &    37.5kpbs                   &    Extremely low      &                                                            Wide reception range, low energy consumption                                                                                                                                                       &  {L}ong distance between base station  and device, sever outdoor-to-indoor signal attenuation due to building walls                                                       \\ \hline
		\textbf{IEEE 802.11ah $\textsuperscript{ \cite{centenaro2015long}}$} &    1km     & 100 Kbps   &   Extremely low    &                                                                      Wide reception range, low energy consumption                                                                                                                                                       & Not thoroughly explored for localization, performance to be seen in indoor environments                            \\ \hline
		\textbf{{Weightless}} 	& {2 km for P,  3 km for N, and 5 km for W}&{100 kbps for N and P, 10 Mbps for W} & {Extremely low} & {Wide reception range, low energy consumption} & {{L}ong distance between base station  and device, sever outdoor-to-indoor signal attenuation due to building walls}  \\ \hline 
	\end{tabular}
	\label{tab:technology}
\end{table*}

\subsection{Acoustic Signal}
The acoustic signal-based localization technology leverages the ubiquitous microphone sensors in smart-phones to capture acoustic signals emitted by sound sources/RNs  and estimate the user location with respect to the RNs. The traditional method used for acoustic-based localization has been the transmission of modulated acoustic signals, containing time stamps or other time related information, which are used by the microphone sensors for ToF estimation~\cite{liu2013guoguo}. In other works, the subtle phase and frequency shift of the Doppler effects experienced in the received acoustic signal by a moving phone have been also used to estimate the relative position and velocity of the phone~\cite{huang2015swadloon}. 

Although acoustic based systems have been shown to achieve high localization accuracy, due to the smart-phone microphone limitations (sampling rate/anti-aliasing filter), only audible band acoustic signals ($<$20KHz) can provide accurate estimations. For this reason, the transmission power should be low enough not to cause sound pollution (i.e., the acoustic signal should be imperceptible to human ear) and advanced signal processing algorithms are needed to improve the low power signal detection at the receiver. Moreover, the need of extra infrastructure (i.e., acoustic sources/reference nodes) and the high update rate (which impacts the device battery), make the acoustic signal not a very popular technology for localization. 

\subsection{Ultrasound}
The ultrasound based localization technology mainly relies on ToF measurements of ultrasound signals ($>$20KHz) and the sound velocity to calculate the distance between a transmitter and a receiver node. It has been shown to provide fine-grained indoor localization accuracy with centimetre level accuracy~\cite{hazas2006broadband,mccarthy2007buzz,priyantha2005cricket}  and track multiple mobile nodes at the same time with high energy efficiency and zero leakage between rooms. Usually, the ultrasound signal transmission is accompanied by an RF pulse to provide the necessary synchronization. However, unlike RF signals, the sound velocity varies significantly when humidity and temperature changes; this is why temperature sensors are usually deployed along with the ultrasound systems to account for these changes~\cite{ijaz2013indoor}. Finally, although complex signal processing algorithms can filter out high levels of environmental noise that can degrade the localization accuracy, a permanent source of noise may still degrade the system performance severely.  
\par Table \ref{tab:technology} provides a summary of different wireless technologies from localization perspective. The maximum range, throughput, power consumption, advantages and disadvantages of using these technologies for localization are summarized. 

\section{{Localization and Internet of Things}}\label{sec:iot}
{The rise of the Internet of Things (IoT) and connecting billions of devices to each other is very attractive for indoor localization and proximity detection. In this section, we provide a primer on IoT and intend to analyze how and if IoT can impact or benefit indoor localization.  } 
\subsection{{Primer on Internet of Things}}
{The Internet of Things (IoT) is based on the fundamental idea of connecting different entities or \textit{`things'} to provide ubiquitous connectivity and enhanced services. This can be achieved by embedding any thing with sensors that can connect to the Internet. It is considered as one of the six \textit{``disruptive civil technologies''} by the US National Intelligence Council (NIC) \cite{zafari2016microlocation,atzori2010internet}. IoT is poised to be fundamental part of the projected 24 billion devices to be connected by the year 2020 \cite{gsmareport} and will generate about \$1.3 trillion in revenue. IoT intends to improve the performance of different systems related to health, marketing, automation, monitoring, parking, transportation, retail, fleet management, security, disaster management, energy efficiency, and smart architecture etc \cite{atzori2010internet}. Indeed, it is expected that by 2025, IoT will be incorporated into food packaging, home  appliances and furniture etc. \cite{iotappliances}. This indicates that the IoT is a technology for the future and have potential for wide-scale adoption. However, augmenting indoor localization into IoT will further enhance the wide range of services that IoT can provide.}
\par {IoT basically can be divided into three components that includes}
\begin{enumerate}
	\item {Sensing/Data Collection: Different sensors or embedded systems connected to the IoT network need to perform a specific task such as sensing temperature, seismic activities, user heart beat, speed of the car etc. The sensors are the fundamental pillars of the IoT systems.  These are usually energy and processing power constrained devices that are not primarily responsible for extensive data processing. Indeed, most of the sensors sense and transmit the data with minimal processing. }
	\item {Data Communication: The data obtained through the sensors need to be communicated to a central entity such as a server, cluster head or intermediate node. This can be achieved through a number of wired or wireless technologies. }
	\item {Data Processing: As IoT would consist of billions of connected devices, the data generated by these devices is going to be huge. Therefore, it requires extensive processing such as data aggregation, compression, feature extraction etc. to provide the user with valuable and related data. This is usually done at a server, which then makes the data available to the user.  Figure \ref{fig:iot2} shows the fundamental building block of an IoT system. }
	
	\begin{figure}
		\centering
		\includegraphics[width=0.4\textwidth]{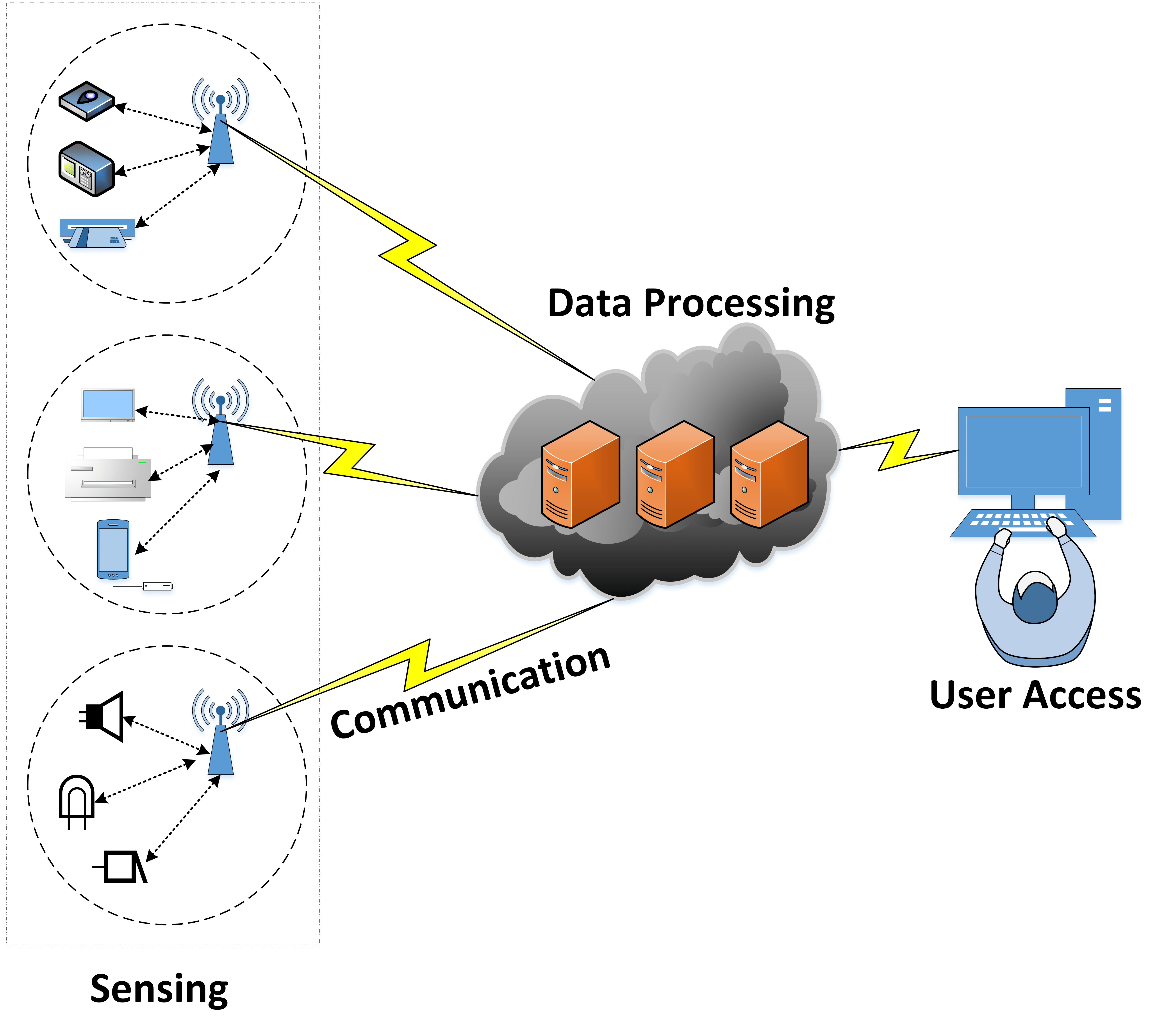}
		\caption{Fundamental building blocks of IoT}
		\protect\label{fig:iot2}
	\end{figure}
	
\end{enumerate}
\par {Cooperation among the connected device of the IoT network is fundamental to its reliable and safe operation. Therefore IoT is going to be highly heterogeneous network that will leverage different communication and connectivity protocols such as cellular, WiFi, Bluetooth, Zigbee, UWB, RFIDs, Long Range Radio (LoRA) \cite{centenaro2015long}, Sigfox \cite{sigfox,centenaro2015long} etc. Such a heterogeneous network poses significant research challenges as the amalgamation of different standards and technologies requires careful deliberation and planning.  IoT is still in its relative infancy and is still evolving. Therefore, it has yet not been thoroughly standardized.  IoT is a single paradigm, but there are a number of potential visions for future IoT that is yet to be agreed upon \cite{atzori2010internet}. There are a number of open research issues in IoT details of which can be found in \cite{atzori2010internet,da2014internet,whitmore2015internet,thoma2012iot,yang2010survey}. }

{By IoT radio we refer not only to the numerous recently emerging IoT communication technologies, such as SigFox, LoRa, WiFi HaLow, Weightless, NB-IoT, etc., but also to a series of existing IoT-enabling wireless communication standards, such as BLE, WiFi, Zigbee, RFID and UWB, which have been used for machine-to-machine communications. This amalgamation of different and diverse technologies allows thousands or millions of devices in IoT ecosystems to connect and communicate in a seamless, yet efficient way. While we discuss the type of services that localization can provide by leveraging the IoT framework in Section VII, below we present and analyse some of the challenges related the indoor localization in IoT ecosystems.} 

\begin{itemize}
	
	
	\item {Privacy and Security: Location privacy is one of the biggest concerns in relation to the use of mobile devices, such as smartphones. Localization in IoT brings another dimension to this problem. The user location can be now correlated with the location of several IoT devices and reveal far more personal information related to the user’s health, mood, behaviour and habits. Even if the user does not carry a mobile device, their location and behaviour can be inferred by processing a sequence of data collected by several and diverse IoT sensors in a close indoor environment (e.g., processing the data from several IoT devices located in different rooms in a smart house can reveal the residents’ everyday habits and behaviour). The appropriate methodology for data collection, processing, encryption and storage to preserve and guarantee the user privacy and anonymity in such environments remains an open and very challenging problem. On the other hand, in an industrial environment, the devices’ location and/or movement may reveal potentially confidential information related to the processes and techniques used by a company to develop a specific product. Random change of the MAC addresses or IDs of the IoT devices, known only by the authorized network, together with appropriate encryption can increase the privacy of the positioning. }
	\par 
	{Safety and security is another concern of paramount importance related to IoT localization. Compromised location information can be very critical for particular IoT applications and services, such as health, structural monitoring, defence, etc. For instance, in structural monitoring, a small change in the distance between two sensors can indicate a severe stability issue of a building. In an industrial setting, the movements of a heavy machinery may depend on the precise calculation of its distance from several IoT sensors. In defence related application, the location of certain IoT devices must remain hidden to the adversary. Although it is evident that localization information tampering or exposure can be proven catastrophic in many of the aforementioned scenarios, in several cases, these devices have been designed considering only their core functionality and not security. }
	\par 
	{Localization does not just create problems, but it can be also proven highly beneficial to the security or several IoT services and applications. Since, many IoT devices are built to last for years, while they are deployed by the end users rather than IoT service professionals, they tend to be placed and forgotten or unintentionally (or intentionally) moved. Knowing the location of these devices enable their recovery when needed. More importantly, the change of a IoT device location may indicate that the device has been compromised (e.g., stolen). In such case, the device may automatically stop harvesting data and delete any data already collected and stored on it. On the other hand, localization may be used for the authentication of new IoT devices, e.g., IoT devices may allow to interact only when they come into close proximity to each other, in order to create trustable services. In other cases, IoT devices may be allowed to communicate only when they locate themselves in “secure” regions to minimize the risk of potential malign transmission overhearing.  }
	\item {Heterogeneity and diversity: IoT ecosystems normally comprise a large number of heterogeneous, diverse and constantly evolving assets and devices. Many of these devices have not been designed/manufactured with localization provisioning, so they can only be located by passively monitoring any data packets they transmit for communication purposes. However, unlike traditional sensor networks, within the same IoT ecosystem, devices may use different wireless air interface protocols (e.g., SigFox, LoRa, WiFi-HaLow, BLE, ZigBee, etc.) and different versions of communication standards. Up to date, the vast majority of existing localization systems are dealing with a single air interface. Developing ubiquitous localization solutions that simultaneously work with a wide variety of wireless protocols is particularly challenging. }
	\par 
	{An alternative solution would be to either modify/enhance the software/firmware of such devices to assist localization or to attach supplementary localization sensors on them. However, in IoT ecosystems (e.g., smart houses, hospitals, industrial environments, etc.) different sets of devices/networks may belong to different service providers, end users or organizations. A service provider that attempts the localization may not necessarily be the owner of the IoT devices/network. In this case, access to these devices to either modify/enhance their software or firmware to assist localization or permission to attach supplementary localization sensors on them, may not be possible.
}\par
		
	{	Moreover, since most of IoT devices have been built to last for several year running on a single battery, they must go through very long sleeping cycles. During these cycles, no packets are transmitted, and the devices are essentially undetectable. This poses significant limitations on how frequently the location of these device can be estimated and updated. Also, when they wake-up, they normally transmit very short packets at low power with minimum repetitions, which results to low signal-to-noise-ration (SNR) and increases the localization error. Depending on the underlying application and energy constraints, the wake-up patterns may vary from device to device, making the design and optimization of unified localization solutions even more challenging.  
	}
\item {Network management and scalability: Interference management is one of the main challenges in IoT networks. When it comes to localization the additional signalling/packet transmission used explicitly for localization may generate significant interference that can reduce the efficiency and disrupt the communication operations not only of the underlying IoT network but also of other conventional networks operating in the same frequency band(s). This is particularly critical in hospital environments where the localization packets may interfere and disrupt the operation of medical equipment. Note that usually in IoT networks the data traffic is primarily generated by a large number of very short and infrequent transmission, therefore the proportion of the localization overhead can be significant (especially if frequent location updates are required) compared to the ongoing data traffic. On the other hand (particularly for devices operating in the ISM bands) interference generated by co-existing (in space, time and frequency) communication networks (e.g., WiFi) may affect the localization process by causing packet collision or wrong measurements (increase the SINR). Moreover, energy spent for localization may decrease the battery life of many IoT devices and this needs to be considered during the initial network panning. Finally, many resource allocation and routing protocols for IoT are based on the actual or relative location of the IoT devices. }
\item
{Most IoT applications (particularly those that require large number of sensors) seek a low per unit cost of their IoT devices, which results in devices with very limited hardware components, such as CPU, memory and battery.  Many of them are purpose made for a particular application (e.g., data collection and transmission) and do not allow for any software or firmware modifications for localization purposes. For others, the low per unit cost limits their computational power, as a result, advanced signal processing algorithms cannot be used for localization on the IoT device side. Many IoT devices are small in size, which makes infeasible the use of multiple antennas or antenna arrays. Embedded antennas like chip and printed circuit board (PCB) are preferred instead, since they have the benefit of fitting into small spaces, but they suffer from low antenna gain and directivity which highly affects the precision of many localization techniques. Another constraint is related to their energy efficiency. IoT devices are built to last on small batteries for periods ranging from few to several years. Additional signal transmissions to assist localization will drain the battery faster and shorten their life expectancy. }
\end{itemize}
\par {Incorporating localization into IoT framework  certainly will help to provide a number of efficient solutions that would improve the overall services provided to the users. However, for that to happen, there is a need for extensive research to find solutions to the aforementioned challenges}

\subsection{{Emerging IoT Technologies}}
{In the following, a number of emerging radio technologies (primarily designed for IoT communication), which can be potentially used for indoor localization will be presented and briefly discussed.}
\subsubsection{{SigFox}}
{Founded in 2009, SigFox is the first \textit{Low Power Wide Area Network} (LPWAN) network operator dedicated to M2M and IoT. 
	Designed to serve a huge number of active devices with low throughput requirements, SigFox operates in the unlicensed ISM radio bands and uses a proprietary Ultra Narrow Band (UNB) radio technology and binary-phase-shift-keying (BPSK) based modulation to offer ultra low data rate ($\sim$100 bits per second) and long range (up to 40 km in open space) robust communication with high reliability and minimal power consumption. 
	By using UNB radio, the noise floor is reduced (compared to classical narrow, medium or wide-band systems); the resulting low reception power sensitivity allows data transmission in highly constrained environments and the ability to successfully serve a huge number of active nodes deployed over a large area with a very small number of base stations.
	Nevertheless, the ultra narrowband nature of SigFox signal makes it susceptible to multipaths and fast fading, which together with the long distance between base stations and devices make the RSS resolution insufficient for localization use. }

\subsubsection{{LoRa}} 
{LoRaWAN is an open Medium Access Control (MAC) protocol which is built on top of the LoRa physical layer (a proprietary radio modulation scheme based on Chirp Spread Spectrum (CSS) technology). LoRaWAN is  designed to provide long-range, low bit-rate communications to large-scale IoT networks and has been already adopted by several commercial (LPWAN) platforms. The uniqueness of LoRa, compare to other IoT technologies, is the use of CSS modulation, a spread spectrum technique where the signal is modulated by frequency varying sinusoidal pulses (known as chirp pulses), which is known to provide resilience against interference, multipath and Doppler effects. 
	These attributes makes CSS an ideal technology for geolocation, particularly for devices moving at high speed, and it was one of the proposed PHYs for the IEEE 802.15.4 standard. However, the bandwidth considered in IEEE 802.15.4 was 80MHz, which is much wider than the typical 125, 250 and 500kHz LoRaWAN bandwidth values. This fact, together with the long-range between the server and the device (i.e., 2-5 km in urban and 15 km in suburban areas), make difficult the multipath resolution and highly reduce the geolocation accuracy of LoRaWAN. Although an ultra-high resolution time-stamp to each received LoRa data packet has been recently introduced by LoRa for TDOA based localization, indoor accuracy cannot be achieved unless additional monitors are deployed in the indoor environment where the devices/users of interest are located {\cite{linklabs}}. {As suggested by \emph{LinkLabs}'s report \cite{linklabs}, using a hybrid approach such as combining GPS-LoRa can achieve better localization accuracy.  }}

\subsubsection{{IEEE 802.11ah}}
{The IEEE 802.11ah, also known as WiFi \emph{HaLow}, is an IEEE standard specification primarily designed for IoT devices and extended range applications. It is based on a MIMO-OFDM physical layer and can operate in multiple transmission modes in the sub-gigahertz license-exempt spectrum, using 1, 2, 4, 8 or 16MHz channel bandwidth. It can operate in multiple transmission modes, from low-rate (starting from 150 Kbps) able to provide whole-house coverage to battery operated IoT devices, such as temperature and moisture sensors; to high-rate (up to 346 Mbps) modes, able to support plug-in devices with power amplifier, such as video security cameras. Its shorter-range network architecture together with the significantly lower propagation loss through free space and walls/obstructions due to its lower operation frequency (compared to LoRa and SigFox), makes IEEE 802.11ah a good candidate IoT technology for indoor localization. Moreover, in contrast with the aforementioned IoT technologies, WiFi HaLow does not require a proprietary hardware and service subscriptions, since off-the-shelf IEEE 802.11ah routers are only needed.}
\subsubsection{{Weightless}}	
{Weightless is a set of open wireless technology standards developed and coordinated by a non-profit group, the Weightless Special Interest Group (SIG), that aim to deliver wireless connectivity for low power, wide area networks specifically designed for the IoT \cite{weightless}. Currently, there are three published Weightless connectivity standards:}
\newline
{
	\textit{Weightless-N} is an uplink only, ultra-narrowband technology (very similar to SigFox) that operates in the sub 1GHz licen{s}e exempt Industrial, Scientific and Medical (ISM) bands. It uses differential BPSK modulation and can deliver 30 kbps to 100 kbps data rate in 3 km range. }
\newline {
	\textit{Weightless-W} exploits the unused ultra-high frequency (UHF) TV white-space (TVWS) spectrum. It can deliver 1 kbps to 10 Mbps data rates in 5 km range (depending on the link budget) and can support several modulation schemes with frequency hopping and a wide range of spreading factors. }
\newline {
	\textit{Weightless-P} is a bi-directional, relatively narrowband (12.5 kHz channels) technology, which is the main focus of the Weightless SIG [ref]. Although it can operate in any frequency band, it is currently defined for operation in the license exempt sub-GHz ISM bands. I{t} is using a non-proprietary physical layer based on Gaussian minimum shift keying (GMSK) and quadrature phase shift keying (QPSK) modulation, which can offer 0.2 kbps to 100 kbps data rate in 2 km range.}
\newline
{
	None of the aforementioned Weightless standards have any built-in localization capability. Although TDOA, RToF or RSS based techniques, described in the previous section, can be used to process the receiving signal for localization purposes, the truth is that the long distance between base stations and IoT devices together with the multipaths, and severe signal attenuation through building walls render them insufficient for localization use. }
\par 

{It becomes evident that although many IoT services will require seamless and ubiquitous indoor/outdoor localization and/or navigation of both static and mobile devices, long-range IoT technologies have not been designed with indoor localization provision. The main issues are (i) the long distance between the base stations and the IoT devices, and (ii) the severe signal attenuation and multi-paths due to the interruption of the outdoor to indoor communication channel by building walls. Even for the `lucky' IoT devices, located within a couple hundred meters from a base station, reasonable indoor localization accuracy is not possible. This is why if accurate indoor localization is required by the underlying application or service, the cooperation between short- and long-range IoT technology will be necessary. The aforementioned long-range technologies can provide low granularity, global location information (e.g., in which building the device is located) while the short-range technologies, such as Bluetooth, Zigbee, WiFi, UWB, RFID, etc., will provide high granularity, local information (e.g., where exactly in the building this device is located). The combination of these two measurement can provide an accurate estimation of the device location within the communication cell. }		
\par{A summary of the aforementioned emerging IoT technologies from a localization perspective has been included in Table \ref{tab:technology}}

\section{Evaluation Framework}
\label{sec:framework}
In this section, we discuss the parameters that we include as part of our evaluation framework.  We believe that for a localization system to have wide-scale adoption, the localization system  must be readily available on user devices, should be cost efficient, energy efficient, have a wide reception range, high localization accuracy, low latency and high scalability. However, it is worth noting that the systems are application dependent and might not be required to satisfy all these metrics. Below we discuss each of them in detail.
\subsection{Availability}
One of the fundamental requirements of indoor localization  is to use a technology that is readily available on the user device and does not require proprietary hardware at the user end. This is important for the wide scale adoption of the technology. UWB based systems have proven to provide 10-20 cm accuracy \cite{zafari2016microlocation}, however, most of the current user devices do not have UWB chip. Similarly, approaches based on {\emph{Synthetic Aperture Radar}} (SAR) might also require additional sensors. Therefore, it is important to obtain localization systems that can work smoothly with widely available devices such as smart phones. Currently, the widely used technology is WiFi, which is readily available on almost all user devices. Similarly visible light and Bluetooth can be used as other viable alternatives. 
\subsection{Cost}
The cost of localization system should not be high. The ideal system should not incur any additional infrastructure cost as well as do not require any high end user device or system that is not widely used. The use of proprietary RNs/hardware can improve the localization accuracy, however it will certainly result in extra cost. While larger companies might be able to afford them, smaller businesses are constrained mostly in terms of such costs. Therefore, we believe that the localization system can easily penetrate the consumer market and be widely adopted by keeping the cost low. 
\subsection{Energy Efficiency}
Energy efficiency is of primary importance from  localization perspective \cite{van2015platform,behboodi2013evaluation}. The goal of localization is to improve the services provided to the users. Any such system that consumes a lot of energy and drains the user device battery might not be widely used. This is because localization is an additional service on top of what the user device is primarily intended for i.e. communication. Therefore, the energy consumption of the localization system should be minimized. This can be achieved by using technology such as BLE that has lower energy consumption or offloading the computational aspect of the localization algorithm to a server or any entity which has access to uninterrupted power supply and has high processing power. The fundamental trade off is between the energy consumption and the latency of the localization system.  Possible factors that can influence the energy consumption of any localization system are
\begin{itemize}
\item Periodicity: The interval or frequency of transmitting the beacon or reference signal for localization significantly affects the energy efficiency, accuracy and latency of the system. The higher the frequency, the higher will be the energy consumption and  accuracy. 
\item Transmission Power: Transmission power also plays a fundamental role in the energy consumption. The higher the signal power, the higher will be the reception range of the localization system and the lower will be the energy efficiency. While transmitting power might not be a major source of concern for MBL systems where the anchor or reference nodes might have access to continuous power and might not rely on the any battery, it is still useful from IoT perspective to optimize the transmission power to obtain a highly accurate but low energy consuming localization system.  Another important factor to consider when dealing with transmission power is the interference. Signals from different reference nodes or the user devices might interfere with each other. 
\item Computational Complexity: Computational complexity of the localization algorithm is also important to take into account. Running a highly complex algorithm on the user device will drain its power source and despite high accuracy, the system might still not be favorable. Therefore, it is important to design algorithms and mechanism which are not computationally complex. As mentioned earlier, the computation complexity can be offloaded to a server at the cost of added delay or latency. 
\end{itemize} 
\subsection{Reception Range}
The reception range of the used technology for localization is also of primary importance in evaluating any system. An industry standard localization system should have a reasonable range to allow better localization in large spaces such as office, hospitals, malls etc. Higher range also means that the number of anchor points or reference nodes required would be low and it will result in cost efficient systems. However, an important aspect to consider is the interference and performance degradation with increase in distance between transmitter and receiver. The choice of the reception range depends on application and the environment in which the localization system is to be used. 
\subsection{Localization/Tracking Accuracy}
One of the most important feature{{s}} of the localization system is the accuracy with which the user/device position is obtained. As mentioned earlier, indoor environments due to presence of obstacles and multipath effects provide a challenging space for the localization systems to operate in. Therefore, it is important for the system to limit the impact of multipath effects and other environment noises to obtain highly accurate systems. This might require extensive signal processing and noise elimination that is a highly challenging task.  The localization system should be able to locate the user or object of interest ideally within 10 cm (known as \textit{microlocation} \cite{zafari2016microlocation})  accuracy. 
\subsection{Latency/Delay}
Real-time localization  requires that the system should be able to report user location and coordinates without any noticeable delay. This means that the system should be able to locate user with a small number of reference signals and should perform complex operations with milliseconds granularity. This is a significant constraint as larger number of reference signals mean obtaining a highly reliable position estimate. However, to report results in real-time, use of extensive signal measurements is not possible. Similarly, the use of complex and time consuming but effective signal processing techniques is also not viable. Therefore, there is a need for optimized signal processing, which should eliminate noise and provide user location with no noticeable delay. 
\subsection{Scalability}
The localization system needs to be scalable\added{{,}} i.e.\added{{,}} it should be able to simultaneously locate or provide services to a large number of users in a large space. 
Scalability is a major challenge to MLB systems when compared with DBL systems as DBL usually happens on the user device, which is not limited by other user devices. However, MBL happens on some monitor or server that is responsible for simultaneously facilitating hundreds of users at a time (in malls, hospitals, sports arenas etc.). 
\par The above factors are important in evaluating any localization system. We do not define any threshold for these metrics as we believe it depends on application and the scale of deployment along with the organization that is utilizing {{the}} localization system. Ideally, there will be a localization system that can satisfy all the above requirements. To the best of our knowledge, there is no such system proposed as of now that satisfies all of these requirements. However, recently some systems have been proposed in the literature that do satisfy majority of the requirements. In the next section, we discuss some of the proposed systems in the literature and evaluate them using our proposed framework. 
\section{Localization Systems}
\label{sec:existing}
In this section, we describe some of the proposed indoor DBL and MBL techniques in the literature. We broadly classify the systems as either device based or monitor based localization and evaluate/compare them from the perspective of energy efficiency, cost, availability, latency, reception range, localization accuracy, and scalability. 
\subsection{Monitor based localization}
We primarily classify the MBL systems based on the wireless technology used. 
\subsubsection{WiFi based MBL}
Bahl et al. present \textit{RADAR} \cite{bahl2000radar}, which is one of the pioneering work that uses RSSI values of the user device to obtain an estimate of the user location \cite{xiong2015pushing}.  During the offline phase,the APs collect RSSI values from the user device that are used to build a radio map. In the online phase, the obtained RSSI values are matched with the offline RSSI values to infer user location. RADAR achieves a median localization accuracy of 2.94 meters (m). Guvenc et al. \cite{guvenc2003enhancements} apply Kalman filter to improve the localization accuracy of a system that uses RSSI from WiFi APs. The authors use RSSI fingerprinting in the offline phase to infer about the position using the RSSI in the online phase. Moving average based system is also compared with Kalman filter to highlight the accuracy attained by a Kalman filter. A median accuracy of 2.5 m is attained. 
\par Vasisht et al. propose \textit{Chronos} \cite{vasisht2016decimeter} that is a single WiFi access point (AP) based MBL system. Chronos uses ToF for accurate localization. The AP receives certain beacon messages from the user device that are used to calculate the ToF. Since accurate localization requires accurate estimation of ToF (order of nanoseconds), Chronos employs the inverse relationship between bandwidth and time to emulate a wideband system. Both the transmitter and receiver hop between different frequency bands of WiFi, resulting in different channel measurements. The obtained information is then combined to obtain an accurate ToF estimate. Once the ToFs are accurately computed at the AP, they are then resolved into distances between each antenna pair on AP and user device (thus both AP and client must be MIMO devices). The measured distances are then used to obtain the 2D locations relative to the AP through an error minimization process (error between measured and expected distances) that is subject to geometric constraints imposed by the antennae's location on each device. While Chronos attains a median accuracy of 0.65 meters, it is not scalable and seems to consume a lot of energy to sweep across different frequencies.
\par  Kotaru et al. \cite{kotaru2015spotfi} propose \textit{SpotFi} that uses CSI and RSSI to obtain an accurate estimate of AoA and ToF, which are used to obtain user location. SpotFi achieves a median localization accuracy of 40 cm using standard WiFi card without the need for any expensive hardware component or fingerprinting. The signals emitted from the user device towards the AP are used to obtain a fine estimate of the AoA using only a limited number of antennas  on the AP. An important observation of SpotFi  is that multipath not only affects the AoA of the signal across various antennas but also the CSI across different WiFi subcarriers (due to different ToF). To account for this, SpotFi uses joint AoA and ToF estimation algorithms by employing the CSI information.   While the system attains a high accuracy  using WiFi APs, it is not suitable for real-time MBL because it cannot calculate position estimate with limited number of signals. 
\par Xiong et al. \cite{xiong2013arraytrack} propose \textit{ArrayTrack}, which relies on accurate AoA calculation at the WiFi AP to estimate user position. It requires comparatively larger number of antennas than SpotFi, however it attains an improved median localization accuracy of 23 cm. ArrayTrack detects the packets at the AP from different mobile user devices, however, it needs to listen to a small number of frames that can be either control frames or data frames (10 samples are used which in time domain accounts to 250 nanoseconds of a packet's samples). Currently it uses short training symbols of the WiFi preamble for detection purposes. ArrayTrack synthesizes independent AoA data from the antenna pairs. For accurate AoA spectrum generation, ArrayTrack uses a modified version of the \textit{Multiple Signal Classification} (MUSIC) algorithm proposed in \cite{schmidt1986multiple}. As MUSIC algorithm without any modification would result in highly distorted AoA spectra, ArrayTrack uses \textit{spatial smoothing} \cite{shan1985spatial} that averages incoming signals across different antennas on the AP. To suppress multipath effects, ArrayTrack relies on the fact that the direct LOS component does not vary drastically across different collected samples while the false peaks or multipath signals do. The obtained AoA spectrum is then used to estimate the user/device location. While ArrayTrack attains a high localization accuracy in real-time and is scalable, the requirement for higher number of antennas is one of its fundamental limitations. Also, it is yet to be seen if the proposed approach can work with commodity off-the-shelf WiFi APs. \par Phaser \cite{gjengset2014phaser} is an extension of the ArrayTrack that works on commodity WiFi and uses AoA for indoor localization. Phaser uses two Intel 5300 802.11 NICs, each with three antennas whereas one antenna is shared between the two NICs resulting in total 5 antennas. To share the antenna, Phaser efficiently synchronizes the two NICs. Phaser achieves a median accuracy of 1-2 meter, which does not satisfy the sub{-}meter accuracy required for indoor localization. ToneTrack \cite{xiong2015tonetrack} uses ToF data to obtain a real-time estimate of user location with a median accuracy of 0.9 meters. ToneTrack combines the ToF data obtained from the channel or frequency hopping of the user device. The channel combination algorithm helps in combining the information from different channels which helps in attaining a fine time resolution suitable for indoor localization. To account for the multipath effects and the absence of LoS paths, ToneTrack uses a novel spectrum identification algorithm that helps in identifying whether the obtained spectrum contains valuable information for localization. Furthermore, by using the triangle inequality, ToneTrack discards those measurements obtained from the WiFi AP that do not have an LoS path to the user device. Numerical results show that ToneTrack can provide fairly accurate and real-time measurements. In terms of the basic principles, Chronos \cite{vasisht2016decimeter} and ToneTrack \cite{xiong2015tonetrack} rely on the same underlying principle of combining information from different channels. ToneTrack is tested with proprietary hardware and it is yet to be seen if it can work with the existing off-the-shelf WiFi cards. {Wang et al. \cite{wang2017csi} proposed a deep learning basd indoor CSI fingerprinting system. The authors use an off-line training phase to train the a deep neural network. In the online phase, probabilistic method is used to estimate the user's location.  An average localization error of as low as 0.9 meters is obtained. Luo et al. \cite{luo2017pallas} present \emph{Pallas} that relies on passively collecting RSSI values at Wi-Fi APs to obtain user location.  The system thrives on passively constructing Wi-Fi database. Pallas initially obtains landmarks present in the Wi-Fi RSS traces, which when combined with the indoor floor plan and location of the Wi-Fi APs is used to map the collected RSS values to indoor pathways. }
\par {Carrera et al. \cite{carrera2018discriminative} proposed a dicriminative learning based approach that combines WiFi fingerprinting with magnetic field readings to achieve room level detection. Then the landmark detection is combined with range based localization models and and graph based discretized system state to refine the localization performance of the system resulting in a localization error as low as $1.44m$.   }

\subsubsection{UWB based MBL}
Ubisense \cite{ubisense} is one of the widely known UWB based MBL system. Ubisense attains an accuracy as high as 15 cm, which is why it is widely used in industries and as a commercial solution. However, cost is one of the leading constraints of Ubisense. Krishnan et al. \cite{krishnan2007uwb} propose a UWB-Infrared based MBL system for robots that can also be used by other entities. UWB readers are placed at known locations and the UWB transmitter attached to the robot transmit UWB pulses, which are then picked up by the UWB readers. TDoA is then used to obtain an estimate of the robot's location. The system accurately tracks a user with root mean square (RMS) error of 15 cm.  Shen et al. \cite{shen2011passive} uses UWB technology for MBL of different objects. The receivers and the transmitter are time synchronized so the system relies on ToF rather than TDoA. The authors assume that the ranging error follows a Gaussian distribution. For MBL, the authors rely on a \textit{Two-Step, Expectation Maximization} (TSEM) based algorithm that attains the Cramer-Rao lower bound for ToF algorithms. The efficiency of the algorithm is verified using simulations that show that error variance is about 30 dB lower than the existing TDoA based approaches. Xu et al. \cite{xu2006position} use TDoA and UWB for locating different blind nodes or users in an indoor setting. The authors take both LoS and NLoS measurements into account and use a TDoA error minimizing algorithm for estimating the location of the user with respect to fixed RNs. 

\subsubsection{Acoustics based MBL}
Mandal et al. \cite{mandal2005beep} present \textit{Beep}, which is an acoustic signal based 3D MBL system. Different acoustic sensors are placed in an indoor environment. The acoustic sensors are connected to a central server through WiFi network. The user device that wants to obtain its position requests position services. Following the request, the device synchronizes itself with sensors through the WiFi network and transmits a predefined acoustic signal. The sensors use the acoustic signal to calculate the ToF and then map into distance.  The distances from all the sensor nodes are then reported to a server that applies 3D multi-lateration for obtaining an estimate of the user location which is then reported to the user using  the WiFi network. The proposed system attains an accuracy of about 0.9m in 95\% of the experiments. While the proposed system is accurate and seems scalable, its energy efficiency, and latency needs to be evaluated.   \par Peng et al. \cite{peng2007beepbeep} propose \textit{BeepBeep} that is an acoustic signal based ranging system. The proposed system can be used for proximity detection system rather than tracking since the authors have only used it for ranging.  The novelty of BeepBeep is that it does not require any proprietary hardware and relies on a software to allow two commodity off-the-shelf devices do ranging to estimate their proximity. Both devices emit special signals called ``beeps'' while simultaneously recording sounds through its microphone. The recording contains the acoustic signal of itself as well as the other device.  The number of samples between the beep signals is counted and the time duration information is exchanged, that is then used for the two way ToF. This results in highly accurate ToF and provides a good estimate of the proximity between the two devices. The reception range of BeepBeep might be a significant problem in large spaces. Furthermore, it is yet to be explored for localization. 
\subsubsection{RFID based MBL}
Ni et al. propose \textit{LANDMARC} \cite{ni2004landmarc} that uses active RFIDs to track the user location. Different RFIDs tags are placed in an indoor environment that serve as RNs.  The object to be tracked such as a user device is equipped with a tracking tag while the  {RN} measures the signal{s} transmitted by the tracking  {tag}. The {RNs are} also equipped with IEEE 802.11b card (Wi-Fi) to  communicate with an MBL server. The {RNs} measure the signal strength of the tracking device  to estimate the device's location.  While LANDMARC is energy efficient and has long range, it has higher tracking latency and has a median accuracy of  1 meter. LANDMARC is also computationally less efficient and requires higher deployment density for achieving improved localization performance. 
 To address these two problems, Jin et al. \cite{jin2006indoor} propose an efficient and more accurate indoor localization mechanism that accounts for the weaknesses of LANDMARC. Rather than relying on the measurements between all the reference tags and the tracking tag, the authors only choose a subset of reference tags based on certain signal strength threshold. This reduces the complexity and improves the localization accuracy. 
\par Wang et al. propose \textit{RF-Compass} \cite{wang2013rf} that utilizes RFIDs on a robot to track the location of different objects which have RFIDs attached to them. RF-Compass relies on a novel space partitioning optimization algorithm to localize the target. The number of RFID tags on the robot reflects the number of space partitions, therefore an increase in the number of RFIDs tags would certainly restrict the target to a small region, hence improving localization accuracy. Furthermore, the increased number of RFIDs on the robot also helps in calculating the device orientation. RF-Compass has a median localization accuracy of 2.76 cm. Wang et al. also propose \textit{PinIt} \cite{wang2013dude} that uses \textit{Multipath Profile} of RFID tags to locate them. PinIt can work efficiently even in the absence of LoS and the presence of different multipath. Reference RFID tags serve as RNs while the multipath profile is built by emulating an antenna array through antenna motion. PinIt works like a proximity detection system that queries the desired RFID tag (attached to the object of interest) and its surrounding tags to locate it. While a median accuracy of 11 cm is attained, PinIt is not widely deployable due to the absence of RFID on majority of the user devices. Furthermore, it can not be used for typical MBL systems.  
\subsubsection{BLE based MBL}
Gonzalez et al. \cite{gonzalez2002bluetooth} present a \textit{Bluetooth Location Network} (BLN) that uses Bluetooth RNs to track the location of a user in an indoor setting. The Bluetooth enabled user device communicates with the Bluetooth RNs, which then transmits the user location information to a master node. The master node is connected to service servers. The BLN system is inspired from typical cellular networks and attains a room level accuracy i.e. it is more suitable for proximity based services. The system has a response time of about 11 seconds which makes it non real-time. 
\par Bruno et al. \cite{bruno2003design} present a Bluetooth based localization system called \textit{Bluetooth Indoor Positioning System} (BIPS). The proposed system has a short range (less than 10m) and is energy efficient. A Bluetooth enabled user device communicates with  fixed Bluetooth RNs that then use a BIPS-server for obtaining an estimate of the user location. All the RNs are interconnected through a network so that they can communicate information to each other. The main tasks of the RNs are a) to act as master nodes and detect the slave (user devices) within its vicinity b) transfer data between the users and the RN.  BIPS can obtain the position of the stationary or slow moving users in an indoor setting. The authors comment on the latency and delay of the system, however, the results do not comment on the localization accuracy. In terms of latency, BIPS system is not feasible for real-time tracking. 
\par Diaz et al. \cite{diaz2010bluepass} present a Bluetooth based indoor MBL system called \textit{Bluepass} that utilizes RSSI values from the user devices to compute the distance between the device and the fixed distributed Bluetooth receivers.  Bluepass consists of a central server, a local server, a Bluetooth detection device and a user device application. User must have the application installed on the device and should login to utilize the MBL system. The local server is for a single map while the central server intends to link different maps. A mean square error (MSE) as low as 2.33m is obtained. 
\par Zafari et al. \cite{faheemthesis} utilize iBeacons for indoor localization services. The RSSI values are collected from different iBeacons on a user device, which forwards the values to a server running different localization algorithms. On the server side, \textit{Particle Filter} (PF), and novel cascaded approaches of using \textit{Kalman Filter-Particle Filter} (KF-PF) and \textit{Particle Filter-Extended Kalman Filter} (PF-EKF) are used to improve the localization accuracy of the system. Experimental results show that on average, PF, KF-PF and PF-EKF obtains accuracy of 1.441 m, 1.03 m and 0.95 m respectively. While the system is energy efficient and accurate, it  incurs significant delay and requires the deployment of iBeacons, which incurs additional cost. {Ayyalasomayajula et al. \cite{ayyalasomayajula2018bloc}} {propose  a CSI based localization systems with BLE technology, which to best of our knowlede is the first work that does so. Since the nature of BLE makes it challenging to use CSI, the authors have proposed BLE-compatible algorithms to address different challenges. An accuracy as large as 86cm is achieved.  Islam et al. \cite{islam2018rethinking} }{  propose a novel multipath profiling algorithm to track any BLE tag in an indoor setting.  The proposed technique has a ranging error of about 2.4m. }

\subsubsection{Ultrasound MBL}
Ashokaraj et al. \cite{ashokaraj2009robust} propose a deterministic approach called \textit{interval analysis} \cite{kieffer2000robust} to use ultrasonic sensors present on a robot for its localization and navigation in a 2-Dimensional (2D) environment. The proposed approach assumes that the map is already available. While methods such as Kalman Filters (KF) or Extended Kalman Filters (EKF) \cite{gustafsson2002particle,gustafsson2010particle,arulampalam2002tutorial} are widely used for  robot localization, the data association step of such methods is highly complex and usually requires linearization. The proposed method bypasses the data association step  and does not require any linearization. The authors provide simulation based results. Furthermore, the paper does not comment on the localization accuracy, latency and scalability of the proposed approach. It is also worth mentioning that the proposed approach relies on the robot's movement and velocity prediction or estimation \footnote{Velocity estimation or prediction is also known as dead-reckoning}. 
\par The \textit{BAT} indoor MBL system proposed in \cite{ward1997new} and experimentally evaluated in \cite{harle2005deploying} uses ultrasonic signals for indoor localization. Due to lower speed of the sounds waves in the air (330 m/s), the accuracy of the localization system significantly improves when compared with other technologies. In a BAT system, the devices to be tracked are provided with proprietary transmitters. The receivers, whose position\added{{s}}  {{are}} fixed and known{{,}} receives the transmitted signal and use it for location estimation of the user. BAT requires the transmitters and receivers to be synchronized. BAT receives an accuracy as high as 3 cm in a 3D space \cite{xiong2015pushing}, however due to the use of ultrasound, its accuracy is very sensitive to the placement of sensors.  Furthermore, it requires a lot of dedicated anchor nodes which is costly. 
\par \textit{Cricket} indoor localization system \cite{priyantha2000cricket} uses a combination of RF and ultrasonic signals for indoor localization. It is complementary to the Bat system as it uses the radio signal only for synchronizing the receivers. Cricket does not require any synchronization between the receiver and transmitter. It achieves an accuracy of 10 cm \cite{xiong2015pushing}, however it requires dedicated hardware and is limited in range due to the use of ultrasonic technology. 
It is worth mentioning here that modern MBL systems highly rely on ubiquitous technologies such WiFi, BLE, and Visible light because they are readily available. However, most of the user devices lack the capability to produce Ultrasonic signals, which is why there are lesser Ultrasound based MBL systems.   

\subsubsection{Visible Light based MBL}
Di Lascio et al. propose \textit{LocaLight} \cite{di2016localight} that uses visible light for MBL. Different RFID sensors are placed on the floor that detect the decrease in the light intensity due to the shadow of the user. The RFID sensors have photodiodes, which is why the system does not rely on any battery or power supply. Under specific settings i.e. the height of the LEDs, the radius of the light zone and height of individuals, the system achieves an accuracy of 50 cm. However, as the RFID sensors need to harvest energy, the system cannot work in real time. Similarly, the system is more suitable for proximity detection than for actual MBL as the system has no information about the user, but detects if any individual is within close vicinity of the light. It is worth mentioning here that visible light based localization systems are  attractive. 
However, it is highly unlikely due to energy and hardware limitations that the user device can transmit visible light for MBL.  
\subsection{Device based Localization}
We primarily classify the DBL systems based on the wireless technology used. Below we discuss some of the existing DBL systems.
\subsubsection{WiFi based DBL}
Lim et al. \cite{lim2005zero} present an RSSI based localization system (the system can also work in an MBL mode) that does not require any offline RSSI fingerprinting phase. WiFi APs, whose position are known a priori serve as the RNs. The APs obtain the RSSI values from other APs that assist in creating an online RSSI map. So the client or the infrastructure measures the RSSI between the client and APs, which is then mapped to distance and used for estimating the user location. While the proposed approach attains a median accuracy as high as 1.76 meters, it requires extra infrastructure (wireless monitors for improving the system performance) that incurs extra cost. Furthermore, the algorithm requires a number of samples to obtain an estimate of the user location, which can incur delay. 
\par Youssef et al. \cite{youssef2005horus} propose \textit{Horus}, which is a WiFi based localization system that relies on RSSI. Horus is a software system on top of the WiFi network infrastructure that relies on fingerprinting to obtain a radio map of the environment during the offline phase. Then using probabilistic technique in the online phase, it provides an estimate of the user location. The offline phase in Horus involves building the radio map, clustering different radio map locations (to reduce complexity) and pre-processing of the signal strength models to account for the spatial and temporal variations in the wireless channel characteristics. While a median accuracy as high as 39  cm is obtained for one of the test beds, Horus relies on fingerprinting and training before it can be used. This makes it highly sensitive to the changes in the environment. 
\par
Kumar et al. \cite{kumar2014accurate} propose Ubicarse, which is a WiFi based localization system that uses a novel formulation of (SAR) on a user device to accurately locate a user within an indoor environment by emulating large antenna arrays. Ubicarse works with user devices that have at least two antennas. The user should rotate the device to emulate SAR as the basic principal is to take snapshots of the wireless channel while the user rotates the device in a certain trajectory. The channel snapshots help in obtaining accurate AoA information that the device can use for accurate localization. The proposed formulation is translation resilient and only relies on angular motion. Ubicarse attains  a median localization accuracy of 39 cm in 3D space. Furthermore, by employing stereo vision algorithms and the camera on the user device, Ubicarse can provide accurate \textit{geotagging} functionality. Ubicarse can provide the global coordinates of the user device while the camera and stereo vision algorithms help to localization different objects. A combination of them provides an accurate global location for different PoIs, which do not have any electronic tag.  While Ubicarse has high localization accuracy, it is evident that emulating large antenna arrays will strain the user device battery. Furthermore, it is not always possible for the user device to have multiple antennas. Also the requirement to twist the device for emulating large antenna arrays means that if the device is lost or in an inaccessible zone, then it is not possible to locate it.\par 
Biehl et al. \cite{biehl2014loco} present \textit{LoCo} that uses WiFi AP and a proprietary framework to  obtain a room level classifier. A classifier is first trained during offline phase using the RSSI values from the WiFi APs using ensemble learning methods called \textit{boosting} \cite{friedman2001elements}.  During the online phase, the collected RSSI values on the user device are then used to estimate the user's probable location. While the proposed framework is energy efficient and will not drain the device battery, the framework can also be implemented on a cloud-as-a-service that the user device can connect to for obtaining its location. The system might be useful for proximity based systems, it cannot be used for many of the localization applications such as indoor navigation or augmented reality (AR). 
\par Bolliger et al. \cite{bolliger2008redpin} propose \textit{Redpin} that relies on fingerprints of RF signals (WiFi, cellular, Bluetooth) to localize users to a room level accuracy which makes it more suitable for PBS. Redpin runs on user mobile phones and follows the basic principles of  \textit{RADAR} \cite{bahl2000radar}.  Rather than utilizing the traditional fingerprinting and training mechanism, Redpin relies on \textit{folksonomy-like} approach that allows the users to train the Redpin system while utilizing its service. Redpin uses collaboration among users and allows users to create and modify  location information. When the user launches the application on his device, the application during its initialization phase called \textit{sniffing}, collects RSSI value of the active cellular (GSM was the system used in the paper) cell, WiFi APs and the ID of all the Bluetooth devices. The collected RSSI values are then forwarded to a server which attempts to estimate user position using the existing RSSI values at the server. If the location is known, it is reported to the user, otherwise the system reports the last known location and continuously obtains measurements to obtain user location. Redpin relies on the \textit{sniffer} module (that runs on the user device) to obtain the RSSI values while the \textit{locator} module (that runs on a server) assists in obtaining the user location estimate using the measurements from the sniffer. While the proposed system eliminates the need for training, it doesn't fulfill the requirements of indoor navigation. Furthermore, the reliance on RSSI may result in lower proximity detection accuracy. The comparison between Redpin and LoCo in \cite{biehl2014loco} shows that LoCo is more accurate and has lower latency than Redpin. This is probably due to the reliance of Redpin on the user collaboration as well as the time taken to complete the map (the author's experiment showed that it took one day to complete the map). \par
Martin et al. \cite{martin2010precise} present an Android application that relies on fingerprinting and RSSI values from the WiFi APs to report to a user, his location on the device. The approach is one of the first approaches to utilize the same device for offline and online phases of the fingerprinting. The authors claim that the proposed system attains an accuracy as high as 1.5m. {Ding et al. \cite{ding2015indoor} present a \emph{Particle Swarm Optimization} (PSO) based approach for location estimation. A novel AP selection algorithm is used to improve the localization accuracy. To refine the accuracy of the system further, Kalman filter is applied to update the initial location estimation. However, the approach seems energy expensive, non-real time and cannot provide sub-meter accuracy. Ding et al. \cite{ding2014efficient} also present a localization system that relies on a novel empirical propagation model. Sparse fingerprints are collected  from different APs which are then used to divide the whole localization space into sub-regions. The proposed propagation model is then used to entirely recover the fingerprint. The fingerprint values are then used to estimate the user location by applying weighted kNN algorithm. However, the proposed approach cannot obtain sub-meter accuracy. }
\subsubsection{UWB based DBL}
Marano et al. \cite{marano2010nlos} provide extensive insight into the performance of UWB radios in an indoor environment. FCC-compliant UWB transceivers are experimentally evaluated to understand the impact of Non-LoS (NLoS) scenarios. The results of the experiments are used to develop a better understanding of the UWB signal propagation in an indoor environment. The extracted features are then combined with regression analysis and machine learning to classify whether an obtained signal is LoS or NLoS. This also helps in reducing the ranging error that arises due to NLoS. 
Ridolfi et al. \cite{ridolfi2016wifi} present a WiFi Ad-hoc system that improves the coverage and scalability of UWB based indoor localization system. The authors propose implementing a UWB based localization system on top of an Ad-Hoc WiFi mesh network. The proposed approach uses the high connectivity and throughput of WiFi, and the accuracy of UWB for a reliable localization system that does not require any existing backbone network. The proposed framework can support 100 users simultaneously with very small roaming delay. \par 
Rabeah et al. \cite{rabeah2016measurement} discuss the problem of blocked LoS in an indoor environment such as a warehouse where the presence of storage racks as well as different impediments greatly influence the presence of LoS. The authors obtain the blocking distribution which can then be used for accurate localization. Yu et al. \cite{yu2004performance} analyze the performance of UWB and ToF based localization using direct-calculation and Davidson-Fletcher-Powell quasi-Newton algorithm. The authors show that both these methods do not rely on any information related to ToF estimation error distribution or variance. 
\subsubsection{Acoustics based DBL}
Guogou \cite{liu2013guoguo} is an acoustic signals-based indoor localization system that requires specific RNs that can transmit acoustic signals, which cannot be perceived by human beings. On the user device side, the microphone uses novel advanced signal processing technique\added{{s}} to receive the acoustic signals from the RN that is then used for localization. Gougou can identify the NLoS signals that helps in improving the overall localization accuracy. The median localization accuracy achieved ranges between 6-25 cm \cite{xiong2015pushing}. However, the reliance on proprietary acoustic RNs, the shorter range of acoustic signals and the effect of sound noise on the performance of Guogou makes it unsuitable for a ubiquitous localization system. Huang et al. \cite{huang2016walkielokie} present \textit{WalkieLokie} that relies on acoustic signals measurement on the user device to calculate the relative position of different entities in the surroundings. WalkieLokie requires the user device to be outside the pockets so that it can receive inaudible acoustic signals from specific RNs or speakers that are primarily intended for marketing and advertisements. While WalkieLokie does not provide the exact location, it is suitable for proximity and relative position based services. Due to the use of acoustic signals, the range of the system is limited to less than 8 m. However, the use of extra RNs can be used to improve the range. To improve the localization accuracy, the authors utilize novel signal processing algorithms and methods to obtain a mean ranging accuracy of 0.63 m. 

\subsubsection{RFID based DBL}
Shirehjini et al. \cite{shirehjini2012rfid} propose an RFID based indoor localization system that relies on a \textit{carpet} of RFID tags and the readers on mobile object to calculate the location and orientation of the mobile device. The proposed system uses low-range passive RFIDs and various other peripherals that help in interpreting the sensor data. The readers on the mobile object reads the information from the RFID tags on the carpet and then uses the information to calculate its position. The proposed system attains an average localization accuracy of 6.5 cm.  Mariotti et al. \cite{mariotti2012wireless} utilize RFID reader within the user's shoes to track the movement of the user in an indoor environment. The RFID reader communicates with the passive RFID tags that are embedded in the floor tiles. The author do not highlight the localization accuracy that their system attained. Willis et al. \cite{willis2004passive} present a passive RFID information grid that can assist blind users in obtaining location and proximity related information. The user shoe is integrated with an RFID reader that can communicate with user device using Bluetooth. An RFID tag  grid, programmed with spatial and ambiance related information, is placed on the ground so that the reader in user shoes can read the position related information and convey it to the blind users.
\par Wang et al. \cite{wang2009implementation} use active RFIDs for localization in an indoor environment. The mobile user device has an RFID reader while fixed RN (RFID tags) are distributed in the environment. The authors use a two step approach. In the first step, the strength of the signal in overlapping spaces is analyzed while in the second phase,  the user's movement pattern is analyzed using the signal strength.  While the proposed localization system is energy efficient, it lacks the accuracy required for certain applications. To improve the accuracy, the number of tags must be increased which can incur extra cost.  Hightower et al. \cite{hightower2001design} propose \textit{SpotON} which is an RFID tags based Ad-Hoc location sensing system. The proposed system relies on the RSSI to obtain the location of different entities in an indoor setting. RFID tags can be installed within a room, with which the tagged users or entities can communicate and obtain their relative location with respect to each other. SpotON can also be used for absolute location, however the absolute position of the RFID tags should be known. 
 
\subsubsection{BLE based DBL}
Zafari et al. \cite{zafari2015enhancing} propose an iBeacon based indoor localization system that uses RSSI. A number of iBeacons are used as RNs that passively transmit  beacon signal{s}. The user device with proprietary iOS application listens to the beacon messages and uses Particle filtering to accurately track the user location with an accuracy as high as 0.97m. The system does not work effectively in real-time due to the inherent \textit{CoreLocation Framework} limitation, which does not allow the user device to report RSSI sooner than 1 second. Furthermore, the use of PF on the user device is not energy efficient and can reduce the device battery life. Kriz et al. \cite{kriz2016improving} combine BLE enabled iBeacon with WiFi based localization system to improve the overall localization accuracy. Initially, the RSSI fingerprints are collected from different RNs and stored in a database. During the online phase, an android application on the user device obtains the RSSI values from different sensors and then estimates user location using the offline values. The use of iBeacons in conjunction with WiFi results in 23\% improvement in localization accuracy and a median accuracy of 0.77 m is obtained. However, the system cannot function in real-time and relies on multiple RSSI values to obtain an accurate estimate. \par 
{Sadowsi et al. \cite{sadowski2018rssi} compare the performance of Wi-Fi, BLE, Zigbee and long-range WAN for indoor localization using RSSI and evaluate their power consumption when they are used by IoT devices.  Experimental results showed that Wi-Fi performed comparatively better than the other evaluated technologies. Zafari et al. \cite{zafari2018icc} present a Particle Filter-Extended Kalman Filter (PFEKF) cascaded algorithm that improves the localization accuracy when compared with using only PF. Sikeridis et al. \cite{sikeridis2018} present a location aware infrastructure that exploits in-facility crowd-sourcing for improving RSSI fingerprinting. The authors develop a probabilistic cell-based model that is obtained using an unsupervised learning algorithm. An average location classification accuracy of 80\% is obtained which is improved to 90\% by using a semi-supervised approach.   } 
{Blasio et al. \cite{blasio2018impact} study the impact of different iBeacon parameters on its performance. Furthermore, to reduce the data collection time, the authors also propose a semi-automatic system. Obreja et al. \cite{obreja2018indoor} also analyze the performance of iBeacons for indoor localization and conclude that beacons are energy efficient.  Ke et al. \cite{ke2018developing} also rely on fingerprinting and filter modifications to improve the localization accuracy of iBeacons. The authors primarily emphasize on the use of iBeacons for smart homes and intelligent systems.   }


\subsubsection{Visible Light based DBL}
Hu et al. \cite{hu2013pharos} present \textit{Pharos}, which is an LED based localization system that requires modification to the existing LEDs. The authors design proprietary system that is connected to the user device which helps in detecting the LED light. Using the RSSI from the LEDs, Pharos calculates the user location with a median accuracy as high as 0.3 m. While the attained accuracy is high, modifying the LED will incur further costs. Similarly, the use of a proprietary detection system (attached to the user device) also will result in higher costs as well as make the system less attractive to the potential users. 
Li et al. propose \textit{Epsilon} \cite{li2014epsilon} that relies on visible light from smart LEDs for localization. The user device is embedded with custom light sensors that can receive the energy transmitted by LEDs. As visible light can cause flicker to human eyes, Epsilon relies on frequency higher than 200 Hz and avoids {lower frequency, since lower frequencies can make people uncomfortable by causing flickering.}  While Epsilon can result in sub-meter accuracy, it requires LoS and at least three RNs (LEDs) to provide user position. Such constraints make Epsilon unsuitable for localization if the user device is in some bag as the LoS requirement will not be satisfied. \par 
Zhang et al. propose \textit{LiTell} \cite{zhang2016litell} that uses fluorescent lights as the RNs and the user device camera as the receiver. The user device camera is converted into a optical sampling device by using image processing algorithms. LiTell relies on the fundamental principle that due to unavoidable manufacturing related reasons, the used RNs have a different characteristic frequency ($>$ 80 KHz) that is imperceptible to human eye but can be detected by the camera on the user device.  LiTell uses this characteristic frequency to differentiate among different RNs and then localize different users based on their proximity to a certain RN. LiTell requires fluorescent lights, which might not be present everywhere, which is why the LiTell based localization system is not readily available. Furthermore, the author do not comment on the localization accuracy that is attained. It would be also interesting to analyse the energy consumed on the user device due to the image processing algorithms. \par Zhang et al. also propose \textit{LiTell2} \cite{zhang2016visible} that collects the light fingerprints of different fluorescent tubes or LED lights and uses photodiodes to collect AoA information for localization. LiTell2 requires a customized AoA sensor that compares the information obtained from two different photodiodes with different field-of-view (FoV)\footnote{The difference in FoV value results in different RSSI values that are then mapped to AoA}. LiTell2 relies on the motion sensors present on the user devices to obtain an accurate estimate of the user location. It is built on top of LiTell \cite{zhang2016litell}, hence it can also work with unmodified fluorescent lights. LiTell2, unlike LiTell, can also work with LEDs as it uses photodiodes. However, just like in the case of LiTell, LiTell2 also needs to be investigated further from localization accuracy and energy consumption perspective. \par
Jung et al. \cite{jung2011tdoa} present an LED based localization system that utilizes the TDoA for tracking user location. The system requires LoS path between the LED transmitters and receivers. The system uses the fact that each LED must have a different frequency. This helps in differentiating among the LED transmitters. Using simulations, the authors show that the proposed system achieves an average localization accuracy of 1.8mm in a 75$m^3$ space. The authors do not comment on the energy efficiency and latency of the proposed system. Therefore, experimentally evaluating it can provide further insights into the system. 
\par 
{Hu et al. \cite{hu2018lightitude} propose a localization system that relies on already existing uneven light distribution in an indoor environment.  The authors  first propose a light intensity model that helps in reconstructing the received light intensity and then a particle filter based module is used to harness user's natural mobility. The proposed system achieves a mean accuracy of $1.93$m in 720$m^2$ office space.   }

\subsection{{Emerging IoT technologies' based localization}}
{Lin et al. \cite{lin2017positioning} discusses the design challenges of positioning support in Narrowband IoT (NB-IoT) and Long Term Evolution (LTE). The paper primarily focuses on the downlink based positioning method called Observed Time Difference of Arrival (OTDOA).  This is in contrast with LoRA which provides uplink based positioning \footnote{The uplink signal from the	device is picked up by different LoRa gateways}. While uplink based positioning has lower device impact, OTDOA has higher scalability. The authors provide an insight into the OTDOA architecture and protocols and also discuss designing the OTDOA positioning reference signals. Simulation results show that the positioning error is more than 50 meters for more than 50\% of the measurements. Sallouha et al. \cite{sallouha2017localization} uses Ultra Narrowband (UNB) long-range IoT networks (Sigfox) for localization. Rather than finding the accurate position, the authors rely on RSSI fingerprinting for classifying the user in particular zone hence the obtained accuracy ranges in tens of meters. Henriksson \cite{RASMUSHENRIKSSONthesis} through simulation shows that the localization accuracy of LoRA increases with increase in the number of RNs. However, the achieved accuracy is in tens of meters. Low Power Wide Area Networks (LPWANs) such as LoRA and Sigfox in its current shape cannot provide high localization accuracy and must be combined with other localization techniques and technologies for higher localization accuracy \cite{raza2017low}. }

\subsection{Miscellaneous Systems}
Other than the widely used above technologies and techniques, there are a number of different systems discussed in the literature as well that relies on Infrared, ambient magnetic field and a number of different technologies. Haverinen et al. \cite{haverinen2009global} present a global indoor localization system that entirely relies on the magnetic field of the environment. The authors argue that the inherent magnetic field of different entities such as walls, doors, windows is unique and can be used as a magnetic signature to identify a location. A magnetometer at the user end is used to sense the magnetic field, that is then matched with offline magnetic field measurements. The approach is very similar to the RSSI fingerprinting based systems. While the authors have not provide a detailed discussion on the accuracy of the system, they have shown that the system can be optimized for an enhanced indoor localization system.
\par   Gozick et al. \cite{gozick2011magnetic} present a detailed discussion on how magnetic maps can be developed for magnetic field based localization.  The internal magnetometer of a mobile phone is used to  collect extensive magnetic field related measurements  at different positions in a building. These measurements can then serve as the reference points for localization in the future.  Riehle et al. \cite{riehle2012indoor} propose a magnetic field based indoor localization system for the blind and visually impaired people. A magnetometer attached to the user body obtains the magnetic field measurements which are then forwarded to the user on his device through Bluetooth. Like the work done in \cite{haverinen2009global}, the system proposed in \cite{riehle2012indoor} also provides 1D localization. Zhang et al. \cite{zhang2015groping} propose \textit{GROPING}, which is a geomagnetic and crowd sourcing based indoor navigation system. GROPING relies on the users to construct the map of any particular floor or building by using the application and measuring the magnetic field at different positions. Once the map is constructed, any  user can then use the constructed map to obtain his location using revised monte-carlo localization. The proposed system is not real-time and the localization accuracy is also more than a meter. 
\par Lu et al. \cite{lu2016dark} propose an image based indoor localization system that relies on thermal imaging to obtain user location. The use of thermal imaging allows the system to work even in the absence of light. To enhance the image quality of thermal images, active transfer learning is used to enhance the classification accuracy. During the classification, the thermal images serve as the targets while color images serve as the sources. Active transfer learning helps in choosing the most relevant sample during the training phase. Experimental results validate the effectiveness of the approach particularly in dark environments. However, the authors do not comment on the latency and localization accuracy of the approach. Furthermore, the experimental results are in terms of detection accuracy rather than the actual location of the user.  
 \par 
 {Distributed localization of wireless sensors has been a topic of research for the last two decades. However with the advent of IoT and the wide-scale use of wireless sensors in IoT paradigm, the use of such distributed algorithms is relevant more than ever.{ } Langendoen et al. \cite{langendoen2003distributed} provides a comparison of three different distributed localization algorithms for WSNs: a) Ad-hoc positioning, b) Robust positioning and c) N-hop multilateration. All these three algorithms work in a three phase structure }
 \begin{enumerate}
 	\item {Determining node-anchor distances: Different nodes share information that help in inferring the distance between anchors and different nodes. This phase primarily revolves around communications among nodes and helps in computing node positions in step 2. }
 	\item {Computing node positions: Different nodes determine their position using the distance estimates to different anchor nodes obtained in phase 1.} 
 	\item {Refining the positions through an iterative process (optional step): This is an optional step in which the position estimates obtained in step 2 are further refined.} 
 \end{enumerate}
 {	Results show that no single algorithm can be declared to be the best as the performance of  algorithms vary with changes in  conditions. But all the algorithms can be improved for higher localization accuracy. These algorithms are also relevant to localization in IoT. If the IoT sensors are treated as anchors and the device to be tracked, such as a smartphone can serve as the node to be tracked. So the above three algorithms can also be applied for IoT based system.}
 \par 
{
 	Savvides et al. \cite{savvides2001dynamic} present \emph{AHLoS} (Ad-Hoc Localization System) that allows different sensor nodes to localize themselves using a number of distributed iterated algorithms. AHLoS requires certain reference nodes (with known locations) and operates in two different phases i.e., ranging and estimation. While ranging allows to estimate the distance between an anchor/RN and a node, estimation would allow the node to locate itself in a 2D or 3D setting using the ranging information. The algorithm proposed is iterative, i.e. the anchor nodes send beacon messages to the neighboring nodes (position unknown) who localize their position using beacon messages. The nodes that localize themselves then act as anchor nodes and send beacon messages to its neighboring nodes whose position is unknown. } 
 {Kumar et al. \cite{nandakumar20183d} present a localization system that consumes only microwatts of power at the mobile terminal  and can attain high localization accuracy. They propose a multi-band backscatter protoype  that works across different frequencies including ISM band. Localization error as low as 145cm is observed at a distance of 60 m.  }
 \par 
Table \ref{tab:solutions} evaluates various proposed localization systems on the basis of metrics we proposed in Section \ref{sec:framework}. The \textit{type} indicates whether it is MBL (M) or DBL (D) system. The \textit{technology} (Tech.) column indicates the type of wireless technology that is employed for localization, while technique highlights what particular  metric is used to obtain user position. \textit{Availability} indicates whether the system can be readily used on the user device i.e. do majority of the user devices around the world have the capability to use the technology? A system will satisfy the cost constraint if it does not require any proprietary hardware or significant modification to the existing infrastructure. Energy efficiency constraint is satisfied only if the user device battery is not significantly drained by the proposed system. As the MBL systems usually do the calculation on APs (which are powered using a direct power supply) or some backend server, so they will mostly be energy efficient unless they require frequent transmissions from the user device that can strain the user device battery. Reception range constraint is satisfied if the reception range is more than 10 meters. Accuracy below 1 meter is considered suitable enough to satisfy the accuracy requirement. Latency must be in order of milliseconds (ms) while we require the system to support multiple device for scalability. The last column proximity basically indicates whether the authors have used their system for proximity based services. \par   Table \ref{tab:accuracy} shows the localization accuracy, advantages and disadvantages of different systems. It is worth mentioning here that the results presented in these papers particularly when it comes to localization accuracy cannot be used for one to one comparison of these systems. This is because of the variations in the environment where the experiments were performed such as the space size, the presence of obstacles, and the number of the people. Furthermore factors such as whether stationary localization\footnote{Where the user device does not move and the measurements for localization are made at specific points in the space with the device not moving} or mobile localization\footnote{Where the user moves with his device and the measurements for localization are made at random points in the space} was carried out also must be taken into account. \textit{EVARILOS} \cite{evarilos} is one such benchmarking project that can assist in comparing different localization projects. Details about the project can be found in \cite{van2013evarilos}. 
\begin{table*}[]
\centering
\caption{Existing systems Proposed in the Literature}
\begin{tabular}{|p{2cm}|c|p{1.1cm}|p{1.05cm}|l|l|l|l|l|l|p{1.1cm}|p{1.1cm}|}
\hline
\multirow{2}{*}{\textbf{System}} & \multirow{2}{*}{\textbf{Type}} & \multirow{2}{*}{\textbf{Tech.}} & \multirow{2}{*}{\textbf{Technique}} & \multicolumn{7}{c|}{\textbf{Evaluation Framework}}                                                                                                                                                                                                & \multirow{2}{*}{\textbf{Proximity}} \\ \cline{5-11}
&                          &                                      &                                 & \textbf{Availability} & \textbf{Cost} & \textbf{\begin{tabular}[c]{@{}l@{}}Energy \\ Efficiency\end{tabular}} & \textbf{\begin{tabular}[c]{@{}l@{}}Reception \\ Range\end{tabular}} & \textbf{Accuracy} & \textbf{Latency} & \textbf{Scalability} & \\ \hline
system in \cite{guvenc2003enhancements}	& \multicolumn{1}{l|}{M}          &        WiFi &  RSSI & $\surd$   &  $\surd$           & $\surd$   &    $\surd$               &   $\times$                &   N/A              &  N/A                  & \multicolumn{1}{l|}{No}               \\ \hline
system in \cite{martin2010precise}	& \multicolumn{1}{l|}{D}          &        WiFi &  RSSI & $\surd$   &  $\surd$           & $\times$   &    $\surd$               &   $\times$                &   $\surd$               &  $\surd$                    & \multicolumn{1}{l|}{No}               \\ \hline
Horus \cite{youssef2005horus}	& \multicolumn{1}{l|}{D}          &        WiFi &  RSSI & $\surd$   &  $\surd$           & $\times$   &    $\surd$               &   $\surd$                &   $\surd$               &  $\surd$                    & \multicolumn{1}{l|}{No}               \\ \hline
RADAR \cite{bahl2000radar}	& \multicolumn{1}{l|}{M}          &        WiFi &  RSSI & $\surd$   &  $\surd$           & $\times$   &    $\surd$               &   $\times$                &   $\surd$               &  $\surd$                    & \multicolumn{1}{l|}{No}               \\ \hline
Loco \cite{biehl2014loco}	& \multicolumn{1}{l|}{D}          &        WiFi &  RSSI & $\surd$   &  $\surd$           & $\times$   &    $\surd$               &   $\times$                &   $\surd$               &  $\surd$                    & \multicolumn{1}{l|}{Yes}               \\ \hline
Redpin \cite{bolliger2008redpin}	& \multicolumn{1}{l|}{D}          &        WiFi, Cellular or Bluetooth &  RSSI & $\surd$   &  $\surd$           & $\surd$   &    $\surd$               &   $\times$                &   $\times$               &  $\surd$                    & \multicolumn{1}{l|}{Yes}               \\ \hline
Ubicarse  \cite{kumar2014accurate}	& \multicolumn{1}{l|}{D}          &        WiFi &  AoA & $\surd$   &  $\surd$           & $\times$   &    $\surd$               &   $\surd$                &   $\times$               &  $\surd$                    & \multicolumn{1}{l|}{Yes}               \\ \hline
Chronos \cite{vasisht2016decimeter}	& \multicolumn{1}{l|}{M}          &        WiFi &  ToF & $\surd$   &  $\surd$           & $\times$   &    $\surd$               &   $\surd$                &   $\times$               &  $\times$  & \multicolumn{1}{l|}{Yes}               \\ \hline	
SpotFi \cite{kotaru2015spotfi}	& \multicolumn{1}{l|}{M}          &        WiFi &  AoA and ToF & $\surd$   &  $\surd$           & $\times$   &    $\surd$               &   $\surd$                &   $\times$               &  $\times$   & \multicolumn{1}{l|}{No}               \\ \hline
ArrayTrack  \cite{xiong2013arraytrack}	& \multicolumn{1}{l|}{M}          &        WiFi &  AoA  & $\surd$   &  $\times$           & $\surd$   &    $\surd$               &   $\surd$                &    $\surd$               & $\surd$   & \multicolumn{1}{l|}{No}               \\ \hline
Phaser \cite{gjengset2014phaser}	& \multicolumn{1}{l|}{M}          &        WiFi &  AoA  & $\surd$   &  $\times$           & $\times$    &    $\surd$               &   $\times$                 &    $\surd$               & $\surd$   & \multicolumn{1}{l|}{No}               \\ \hline
BAT \cite{xiong2015pushing,harle2005deploying}& \multicolumn{1}{l|}{M}          &        Ultrasound &  ToF  & $\times$   &  $\times$           & $\surd$ &    $\times$          &   $\surd$                &    $\surd$               & $\surd$   & \multicolumn{1}{l|}{No}               \\ \hline
Cricket \cite{priyantha2000cricket}	& \multicolumn{1}{l|}{D}          &       Ultrasound \& RF & ToF   & $\times$   &  $\times$           & $\times$    &    $\surd$               &   $\surd$                 &    $\surd$               & $\surd$   & \multicolumn{1}{l|}{No}               \\ \hline	
Guoguo \cite{liu2013guoguo}	& \multicolumn{1}{l|}{D}          &       Acoustic Signals &  ToF  & $\surd$   &  $\times$           & $\times$    &    $\times$               &   $\surd$                 &   $\times$                & $\surd$   & \multicolumn{1}{l|}{No}               \\ \hline		
WalkieLokie \cite{huang2016walkielokie}	& \multicolumn{1}{l|}{D}          &       Acoustic Signals &  N/A  & $\surd$   &  $\times$           & $\times$    &    $\times$               &   $\surd$                 &   N/A                & $\surd$   & \multicolumn{1}{l|}{Yes}               \\ \hline	
Beep \cite{mandal2005beep}	& \multicolumn{1}{l|}{M}          &       Acoustic Signals &  ToF  & $\surd$   &  $\times$           & $\times$    &    $\surd$               &   $\surd$                 &   $\times$                & $\surd$   & \multicolumn{1}{l|}{No}               \\ \hline			
iBeacon based system in \cite{zafari2015enhancing}	& \multicolumn{1}{l|}{D}          &       Bluetooth &  RSSI  & $\surd$   &  $\times$           & $\times$    &    $\surd$              &    $\times$                 &   $\times$                & $\surd$   & \multicolumn{1}{l|}{No}               \\ \hline		
iBeacon based system in \cite{faheemthesis}	& \multicolumn{1}{l|}{M}          &       Bluetooth &  RSSI  & $\surd$   &  $\times$           & $\surd$    &    $\surd$              &    $\surd$                 &   $\times$                & $\times$   & \multicolumn{1}{l|}{Yes} 
 \\ \hline	
Bluepass \cite{diaz2010bluepass}	& \multicolumn{1}{l|}{M}          &       Bluetooth &  RSSI  & $\surd$   &  $\times$           & $\surd$    &    $\surd$              &    $\times$                 &   $\surd$                & $\times$   & \multicolumn{1}{l|}{No}               \\ \hline	              
ToneTrack \cite{xiong2015tonetrack}	& \multicolumn{1}{l|}{M}          &       WiFi &  TDoA & $\surd$   & $\times$         & $\surd$    &    $\surd$              &    $\surd$                &   $\surd$                & $\surd$   & \multicolumn{1}{l|}{No}               \\ \hline	
RF-Compass \cite{wang2013rf}	& \multicolumn{1}{l|}{M}          &       RFID &  MP\footnote{Multipath Profile} & $\times$   & $\surd$         & $\surd$    &    $\surd$              &    $\surd$                &   $\surd$                & $\surd$   & \multicolumn{1}{l|}{No}               \\ \hline	
PinIt \cite{wang2013dude}	& \multicolumn{1}{l|}{M}          &       RFID &  MP\footnote{Multipath Profile} & $\times$   & $\surd$         & $\surd$    &    $\surd$              &    $\surd$                &   $\surd$                & $\surd$   & \multicolumn{1}{l|}{Yes}               \\ \hline	
LANDMARC \cite{ni2004landmarc}	& \multicolumn{1}{l|}{M}          &       RFID & RSSI & $\times$   & $\surd$         & $\surd$    &    $\surd$              &    $\times$                &   $\times$                & $\surd$   & \multicolumn{1}{l|}{No}               \\ \hline	
LocaLight \cite{di2016localight}	& \multicolumn{1}{l|}{M}          &       Visible Light &  N/A & $\surd$   & $\surd$        & $\surd$    &    $\times$              &    $\surd$                &   $\times$                & $\times$   & \multicolumn{1}{l|}{No}               \\ \hline	
LiTell \cite{zhang2016litell}	& \multicolumn{1}{l|}{D}          &       Visible Light &  N/A & $\times$   & $\surd$        & $\times$    &    $\times$              &    N/A                &   $\surd$                & $\surd$   & \multicolumn{1}{l|}{No}               \\ \hline	
LiTell2 \cite{zhang2016visible}	& \multicolumn{1}{l|}{D}          &       Visible Light &  N/A & $\times$   & $\surd$        & $\times$    &    $\times$              &    N/A                &   $\surd$                & $\surd$   & \multicolumn{1}{l|}{No}               \\ \hline	
Pharos \cite{hu2013pharos}	& \multicolumn{1}{l|}{D}          &       Visible Light &  RSS & $\times$   & $\times$        & $\times$    &    $\times$              &    $\surd$                &   N/A                & $\surd$   & \multicolumn{1}{l|}{No}               \\ \hline	
System in \cite{jung2011tdoa}	& \multicolumn{1}{l|}{D}          &       Visible Light &  TDoA & $\times$   & $\times$        & $\times$    &    $\times$              &    $\surd$                &   N/A                & $\surd$   & \multicolumn{1}{l|}{No}               \\ \hline	
System in \cite{kriz2016improving}	& \multicolumn{1}{l|}{D}          &      WiFi \& iBeacons &  RSSI & $\surd$   & $\times$        & $\times$    &    $\surd$              &    $\surd$               &   $\times$                & $\surd$   & \multicolumn{1}{l|}{No}               \\ \hline	
System in \cite{lim2005zero}	& \multicolumn{1}{l|}{M/D}          &      WiFi &  RSSI & $\surd$   & $\times$        & $\times$    &    $\surd$              &    $\times$               &   $\times$                & $\surd$   & \multicolumn{1}{l|}{No}               \\ \hline	
System in \cite{shirehjini2012rfid}	& \multicolumn{1}{l|}{D}          &      RFID &  N/A & $\times$    & $\times$        & $\surd$    &    $\surd$              &    $\times$               &   $\times$                & $\surd$   & \multicolumn{1}{l|}{No}               \\ \hline	
\end{tabular}
\label{tab:solutions}
\end{table*}
\begin{table*}[]
	\footnotesize
	\raggedleft
	\caption{Claimed localization accuracy of different localization systems}
	\vspace{-0.09in}
	\begin{tabular}{|l|l|p{6.2cm}|p{6cm}|}
		\hline
\textbf{System} & \textbf{Accuracy} & \textbf{Advantages} & \textbf{Disadvantages} \\ \hline
System in  \cite{guvenc2003enhancements}	&  2.4 m median   & {R}eadily available, does not require extra hardware  &  {L}ow accuracy, no information provided about latency and scalability of the system in the paper        \\ \hline
Horus \cite{youssef2005horus}	&  39 cm median   & {A}ccurate, scalable and readily available & {R}equires fingerprinting, may not be energy efficient.       \\ \hline
RADAR \cite{bahl2000radar}	&  2.94 m median   & One of the pioneering fingerprinting based work  & {L}ess accurate, energy inefficient      \\ \hline
Ubicarse \cite{kumar2014accurate}	&  39 cm median   & Novel formulation of synthetic aperture radar that attains high accuracy, does not require any fingerprinting, incurs no extra hardware cost   & Requires the user to twist the device for localization, devices must have two antennas, not energy efficient,  might affect the throughput offered by AP to other users.    \\ \hline
SpotFi \cite{kotaru2015spotfi}	&      40 cm median & Highly accurate, incurs no extra cost  & {M}ight not be scalable,  can drain the device battery, is not suitable for real time localization, might affect the throughput offered by AP to other users.    \\ \hline
Chronos \cite{vasisht2016decimeter}	&   65 cm median & {H}igh accuracy, and does not require extra hardware, only requires one AP for localization  &  {M}ight affect the throughput offered by AP to other users, can affect the battery of the user device due to sweeping across different frequencies for accurate ToF calculations  \\ \hline
ArrayTrack \cite{xiong2013arraytrack}	& 23 cm median  & {R}eal time and accurate MBL system,  & {R}equires some modifications to the AP that can incur extra cost, also might affect the performance of the APs.   \\ \hline
Phaser \cite{gjengset2014phaser}	& 1-2 m median   & {S}uitable reception range, real-time and scalable &  {O}ver a meter accuracy, not energy efficient an requires modifications to the APs for.    \\ \hline
BAT \cite{xiong2015pushing,harle2005deploying}	& 4 cm median& {H}ighly accurate, one of the pioneering work & {R}equires extra hardware which will incur further cost, Ultrasonic systems are not widely used.                  \\ \hline
Cricket \cite{priyantha2000cricket}	&   10 cm median  & {H}ighly accurate and scalable & {R}equires extra hardware which will incur further cost, Ultrasonic systems are not widely used.                \\ \hline
Guoguo \cite{liu2013guoguo}	&   6-25 cm median  & {H}igh accuracy  &  requires extra RNs, cannot work in high sound pollution, not real-time             \\ \hline
System in \cite{zafari2015enhancing}	&   97 cm highest  & {G}ood accuracy and readily available on the user device & {N}ot real-time, requires extra hardware     \\ \hline
Beep \cite{mandal2005beep}	&   0.9 m 95\%  & {A}ccuracy and privacy  &  {R}equires extra RNs, cannot work in high sound pollution             \\ \hline
System in \cite{zafari2015enhancing}	&   97 cm highest  & {G}ood accuracy and readily available on the user device & {N}ot real-time, requires extra hardware     \\ \hline
System in \cite{faheemthesis}	&   95 cm average  &{G}ood accuracy and readily available on the user device & {N}ot real-time, requires extra hardware    \\ \hline
ToneTrack \cite{xiong2015tonetrack}	&   90 cm median   & {R}eal-time, accurate, and energy efficient   & {W}ill not work if the device is not transmitting\\ \hline
RF-Compass \cite{wang2013rf}	&   2.76 cm median   & {H}ighly accurate, provides device orientation as well  &  {N}ot real-time, performance is tested in  very small place  \\ \hline
LANDMARC \cite{ni2004landmarc}&   1 m median   & {E}nergy efficient, and fairly accurate  &   {C}omputationally less efficient, requires high deployment density for its performance, tested in a small scale  \\ \hline
PinIt \cite{wang2013dude}	&   11 cm median   & {H}igh accuracy, energy efficient, reasonable range & {N}ot readily available on {{a}} majority of the user devices, not suitable for typical MBL systems    \\ \hline
LocaLight \cite{di2016localight}	&  50 cm    & {H}igh accuracy & {R}equires specific type of lighting, cannot work in NLoS   \\ \hline
LiTell \cite{zhang2016litell}	&  N/A   & {R}eal-time and low cost & {N}eed LoS   \\ \hline
LiTell2 \cite{zhang2016visible}	&  N/A   &  {R}eal-time and low cost & {N}eed LoS   \\ \hline
Pharos \cite{hu2013pharos}	&  0.3 m median  &  {H}igh accuracy, relies on lighting & {R}equires LoS, might not be real-time     \\ \hline
System in \cite{jung2011tdoa}	&  1.8 mm average  & {H}ighly accurate  &  The authors do not comment on
the energy efficiency and real time nature of the proposed
system   \\ \hline
System in  \cite{kriz2016improving}	&  0.77 m median   & {F}airly accurate & {N}ot real-time, requires the user to be stationary   \\ \hline
System in \cite{lim2005zero}	&  1.76 m median   &  {C}an be used for both MBL and DBL & {R}equires extra hardware, the accuracy is over 1m, not real-time \\ \hline	
System in  \cite{ubisense}	&  0.15 m maximum   & {H}ighly accurate, widely used in industries & {H}igh cost  \\ \hline
System in \cite{krishnan2004system}	&  0.15 m RMS  & {H}ighly accurate & {I}ncurs extra cost, requires extra hardware, not widely available on the user devices  \\ \hline
System in \cite{martin2010precise}	&  1.5 m highest  & {O}ne of the first papers to use same device for offline and online phase & {R}equires fingerprinting \\ \hline
WalkieLokie \cite{huang2016walkielokie}	& 0.63 m mean   & {H}igh accuracy  & {L}imited range, can be affected by sound pollution \\ \hline
LoCo \cite{biehl2014loco}	& 0.944 proximity detection   & {S}uitable for proximity based services & {L}ow accuracy and requires fingerprinting  \\ \hline
Redpin \cite{bolliger2008redpin}	& 0.947 proximity detection   & {S}uitable for proximity based services & {L}ow accuracy and requires fingerprinting   \\ \hline
System in \cite{shirehjini2012rfid} & 6.5 cm proximity detection   & {H}igh accuracy, energy efficient & {T}ested in a very small area  \\ \hline
\end{tabular}
\label{tab:accuracy}
\end{table*}
\section{Applications of Localization}
\label{sec:applications}

{Traditional indoor location-based services have been mainly associated with people positioning and tracking (i.e., by exploiting the wireless signal transmitted from their personal devices) and the use of dedicated wireless sensor networks for asset tracking. Such traditional services, together with several innovative and constantly emerging applications, have become integral part of the wider IoT paradigm. Novel IoT applications are driven by: }
\begin{enumerate}
	\item {	The wide proliferation of IoT which is becoming ubiquitous in almost every modern indoor environment (e.g., smart houses, hospitals, schools, molls, factories). }
	\item {The technological capability and diversity of IoT devices, ranging from low-cost sensors to sophisticated smart devices which can collect data and interact with the environment and the end-user in myriad different ways.}
	\item {	The amalgamation of different IoT technologies (i.e., SigFox, LoRa, WiFi HaLow, Weightless, NB-IoT, etc.)  and other relevant IoT-enabling wireless standards such as BLE, WiFi, Zigbee, RFID and UWB, which allow all these devices to connect and communicate in a seamless, yet efficient way.}
	\item {The constantly increasing market demand for new commercial products and improvement of user experience. }
\end{enumerate}
{Localization and tracking can either be the primary purpose of several IoT devices and network deployments (i.e., dedicated sensors to locate the absolute or relative positions of people, animals or objects), or a value-added service to many other IoT systems (i.e., exploit the wireless signal used for the communication of IoT devices to estimate and add location information to the collected by a device data). For these reasons, localization applications have recently seen a drastic increase in use around the world. They vary from marketing and customer assistance, to health services, disaster management and recovery, security, and asset management and tracking. This section provides a brief overview of these applications}
\subsection{Contextual Aware Location based Marketing {and Customer Assistance}}
Marketing is a fundamental part of any business, as it allows to improve the image of the brand and the product and helps in attracting more customers that ultimately leads to higher sales and profits. Traditional marketing is carried out through different advertisements on televisions, mails, billboards, emails, phones etc. However, they are not optimized sources of advertisements as such advertisements do not usually take the customer location as well as context such as age, ethnicity, gender etc. into account.
\par Contextual-aware location based marketing is fundamentally a revolutionary idea in the world of marketing that is poised to improve the sales and profits. Rather than spamming customers with irrelevant product advertisements, such marketing would allow the business owners with the opportunity to only send relevant advertisements and notifications. For example, any customer `x' who is primarily interested in sports equipment would be sent advertisements/coupons relevant to his/her interest based on his proximity or location in the store. While the location can be obtained through indoor localization systems, the context can be obtained using the historical data (customer's past visit data for inference). While the idea is in its relevant infancy, the rise of \textit{big data analytics} and IoT is going to fuel its adoption. 
Museum 2.0 is one other such novel concept that intends to improve the visitor satisfaction level by enhancing the overall experience in a museum. Localization is a fundamental part of Museum 2.0 in which the user location and interest is taken into account to provide relevant information to the users. The museum localization system can make the artifacts interactive by playing videos or sounds when any user approaches any particular exhibition piece. Furthermore, the museum can alert the visitor about a exhibition within the museum taking the user interest into account. Through localization, the user can then be navigated to a particular exhibition.  
\par Similarly, other environments such as libraries and airports can also greatly benefit from location based services. In libraries, the visitors can find a specific book and the location of the book using localization. Similarly, the library can also provide the student with relevant information based on the location. In airports, localization can allow the customers to find their respective boarding gates or terminals without any hassle and wastage of time. Major airports such as John F. Kennedy (JFK) in New York, Heathrow London, Miami International and many more have started using \textit{iBeacons} to provide proximity based services to the travelers and improve overall customer experience \cite{ibeaconairport}. In fact, Japan airlines  uses MBL to obtain the location of its staff and accordingly assign tasks \cite{ibeaconairport} in Tokyo Haneda Airport. 

\subsection{Health Services}
Health sector can greatly benefit from indoor localization  as it can help save valuable lives. It can help both the hospital staff, the patients as well as the visitors{.} 
 If a patient needs medical assistance, the current protocol requires broadcasting the message or paging a specific doctor or staff member who may not be in vicinity of the patient. The delay in the arrival of the staff might even cause the death of the patient. Similarly, broadcasting the message will cause other staff members to receive irrelevant messages. A location based solution would allow to track the position of the staff members. In case of emergency, the localization system would find the staff member who is in close vicinity and has the necessary qualification to handle the emergency situation. This will avoid the aforementioned delay as well as not spam the other staff members. Indoor localization can also allow the doctors to track various patients and track their mobility to ensure patient safety. Visitors who intend to visit patients can find their destination using a  localization system without any hassle. 
\subsection{Disaster management and recovery}
Technology can facilitate disaster management and assist in recovery following any natural disaster (such as tornado, earthquakes, storms and flood etc.)  or human caused disasters (terrorist attacks etc.). Localization can also help in efficient disaster management and expedite the recovery process. One of the fundamental challenges of disasters is usually obtaining information about human beings, whether they are safe or not and what is their location in the disaster affected area. Localization can help in such scenarios by providing the accurate location of the missing individuals and providing them with medical help in extreme scenarios such as the user being stuck in a rubble after an earthquake. Similarly, in the event of a fire or any other calamity in an indoor environment, the rescue team can obtain the user locations through the localization system that can be then used for targeted operation in the affected location.  
\subsection{Security}
Localization can greatly improve security conditions around the world. User mobility patterns and interaction can be used to identify possible threats that might pose security risks. Similarly, in battlefield or war zones, the military can track its assets and troops through a localization system that will improve the overall operation and increase the chances of successful operation. The soldier on ground can also benefit from a robust localization system to navigate in areas not known to them. This is a strategic advantage as the soldiers can pay attention to their operation and not worry about the paths to take for moving forward. Using localization, the central command can design better strategies and plans, which they can then provide to the soldiers on the ground. 

\subsection{Asset Management and tracking}
Asset management can fundamentally benefit from tracking as it would allow different businesses to track the location of their assets. It will also allow for better inventory management and optimized operation management. While asset management and tracking has been extensively discussed in literature \cite{slavin2015fixed,brown2016method,stevens2013wireless,emanuel2013apparatus,balakrishnan2012efficient,srinivasan2013rfid,jeong2014fully,chakravarty2015system}, we believe that the advent of IoT along with accurate indoor localization system will revolutionize asset management and tracking. Use of novel energy efficient techniques and algorithms will eliminate the need for expensive proprietary hardware that is currently used in the industry and different firms. 

\par The aforementioned applications show that localization can provide us with efficient and effective services, motive behind which is to help the users and customers. In future, we expect a wide range of other services and applications that would be possible due to indoor localization. 
\section{Challenges}
\label{sec:challenges}
In this section, we highlight some of the significant challenges that indoor localization and its adoption faces. 

\subsection{Multipath Effects and Noise}
A fundamental challenge of indoor localization is the presence of multipath effects. Due to the inherent nature of the signals, they can be reflected, refracted and diffracted of the walls, metals, and in some cases even human beings. This drastically affects the behavior of the signals. Approaches such as RSSI, ToF, TDoA, AoA rely on these signals from the RN or the user device to estimate the user location. However, in presence of multipath effects, it is highly unlikely to obtain a single signal. The receiver usually receives a number of different phase delayed and power attenuated versions of the same signal, which makes it challenging to obtain the direct LoS signal and estimate the actual distance between the transmitter and receiver.  This has significant consequence on indoor localization particularly the accuracy. To obtain accurate estimate of the location, there is a need for complex signal processing techniques that can identify the LoS signal (if there is any) and minimize/eliminate the effects of multipath signals. While recently literature has proposed some novel and effective multipath and noise suppressing algorithms, their adoption and utilization on wide scale seems highly unlikely as such algorithms are complex and primarily feasible for MBL (as MBL is at RN which usually has higher processing capability and is not constrained in terms of power). However, for DBL, such complex algorithms might not be useful since most of the user devices lack the energy and  processing power to run such algorithms. Therefore, there is a need for optimized, energy efficient and effective multipath and noise suppressing algorithms that can assist in employing the signals for accurate localization
\subsection{Radio Environment}
Indoor localization is highly dependent on the characteristics of the indoor environment. The performance of the system highly varies with the variation in dynamics of the environment such as what are the walls and ceilings made up of, how are different entities which act as obstacles placed and how many people are there in the indoor space. All these factors must be taken into account when designing any accurate localization system. Most of the existing systems are tested in controlled environment and they do not necessarily replicate the characteristics of a real world indoor environment. It is assumed in most of the proposed systems that there must be at least one LOS path between the user and the RNs. However, in big malls or small offices, it is highly likely that there will be no LOS path between the user device and the RNs. Therefore, there is a need to accurately model the characteristics of the indoor environment. The model must take into account all the variations in the environment particularly the impact of the human beings during peak and off-peak hours of operation.  
\begin{table*}[]
	\centering
	\caption{Challenges of indoor localization and suggested solutions}
	\begin{tabular}{|p{4cm}|p{6.5cm}|p{6.5cm}|}
		\hline
		\textbf{Challenge} & \textbf{Description} & \textbf{Suggested Solution} \\ \hline
		\textbf{Multipath \& Noise}	&  The presence of obstacles and different interfering signals can affect the performance of the indoor localization systems                    &    Utilize schemes that are energy efficient, less complex and robust enough to minimize the adverse effects of multipath and noise. Chirp spread spectrum is one of the probably techniques robust to multipath and noise.                \\ \hline 
		\textbf{Environment Dynamics}	& The change in the environment where the localization system is used  such as the number of people, the presence of different equipment such as cupboards, shelves etc. makes it much more challenging to accurate{ly} obtain user position. & Do not rely on approaches such as fingerprinting, but rather use the improved signal processing and superior computing power of the user devices and servers to obtain metrics such as CSI that are not highly influenced by environment dynamics. Furthermore, always design the system keeping in mind the worst case.            \\ \hline
		\textbf{Energy Efficiency}	&   Real-time and highly accurate localization  might drain user device and RN's power. However, for wide-scale adoption of any indoor localization system, the system must be highly energy efficient.   & Use less complex and energy efficient algorithms. Offloading complex algorithms to some servers or cloud based platform is also more energy efficient when compared with using user device.                  \\ \hline
		\textbf{Privacy and Security}	&  User location is a sensitive information that a lot of users are not willing to share. This is one of the reasons that indoor localization has not yet been adopted on a wide scale.  & The localization service providers should guarantee the users that the information will only be used for the agreed upon purposes and will not be shared with any other entity. Similarly, there is a need for new laws and legislations that guarantees {the user privacy will be protected. }                    \\ \hline
		\textbf{Cost}	&  The use of extra hardware or proprietary systems for indoor localization  is a major hurdle to its adoption particularly when it comes to small business which might not be able to afford them. & use existing architecture, possibly WiFi, for providing localization services without requiring any extra hardware.                     \\ \hline
		\textbf{Lack of Standardization}	& Currently there is no standard that can govern indoor localization research. Therefore, there are a number of orthogonal solutions.                      &   There is a need to define a standard for future localization. The standard should take into account different applications and requirements and accordingly set the benchmark or minimum requirements as done by 3gpp \cite{3gpp}.                 \\ \hline
		\textbf{Adverse affects on the used technology}	&  Localization relies on different wireless technologies primary purpose of which is to connect users and provide improved throughput. As localization  is secondary purpose, it can impact the primary purpose of connecting users as highlighted in \cite{vasisht2016decimeter}                    & Utilize mechanisms and techniques that can provide accurate localization  with minimal effect on the primary purpose of wireless technologies. Localization should be orthogonal to the primary purpose of these technologies.                   \\ \hline
		\textbf{Handovers}	&   The limited range as well as heterogeneity of wireless technologies makes it very challenging to obtain a real-time and reliable system.                   &  Novel, robust and energy efficient hand-over (both vertical and horizontal) mechanisms must be researched and utilized for indoor localization.                 \\ \hline
	\end{tabular}
	\label{tab:challenges}
\end{table*}

\subsection{Energy Efficiency}
Energy efficiency of the localization systems is very important for their ubiquitous adoption. As of now, most of the existing localization systems comparatively use higher energy to provide higher accuracy and better range. Particularly for localization systems, it is extremely challenging to obtain high accuracy without straining the device battery. This is because for improved localization performance, user device has to periodically listen to specific beacon message or signals. This requires the device to actively monitor the wireless channel and pick up different signals. While this is feasible performance wise, it is not ideal in terms of energy efficiency. As localization is the secondary task of most of user devices, the drainage of device battery can lead to user dissatisfaction.  Therefore, there is a need to optimize the energy consumption of the localization  system. While the current research focuses on improved localization performance in terms of accuracy, in future there is a need for also optimizing the energy consumption of the systems. Using highly effective noise suppressing but less complex localization algorithms would help keep the energy consumption cost low. In case of DBL, the user device can offload the computational aspect of the localization to some local or cloud based server that usually has high processing power and continuous power supply. In such cases, latency or the response time also needs to be optimized as the goal is to provide real-time location updates to the user. 
\subsection{Privacy and Security}
The fundamental challenge to the adoption of wide scale localization services is privacy. Most of the subscribers or users are not willing to share data related to their location. This is because user location is  {very sensitive information that}  can jeopardize user privacy and security. Currently, the existing localization  systems do not taken into account the privacy concerns and are primarily concerned with accurate and effective indoor localization. However, with the ever increasing Cyber-security challenges and the lack of an underlying privacy mechanism for indoor localization, privacy is a major challenge that the researchers have to address. How do we guarantee, that a user who uses localization services will not have privacy issues and the user data will be kept secure, confidential and only used for specific purposes such as targeted marketing etc.$?$ Furthermore, how can the user trust the system and the localization service provider? These are fundamental questions that needs to be addressed to address the privacy issues of localization.  The deficit of trust between the users and service providers and the security challenges that can arise as a result of privacy breach needs to be thoroughly tackled in order to allow for localization services to flourish. Also, the system needs to authenticate that the new user who wants to use localization services is not a malicious node but indeed a customer who intends to benefit from the provided services. If the authentication mechanism is weak, a malicious node can infiltrate the system and carry out a systematic attack against the localization system that will certainly affect the overall performance of the system. Novel optimized security and privacy protection mechanisms need to be put in place to guarantee user safety and improved services. Using the traditional complex and processing extensive centralized or distributed key based systems will not work with the energy constrained devices. There is a need for a privacy and security mechanism, that is secure, energy efficient and does not require high computing power. While these constraints are orthogonal to each other and requires making {a} trade-off between the processing complexity and privacy and security, an optimal trade-off point can likely be reached. Another possible solution is to design the system as a location support system rather than a location tracking system \cite{priyantha2000cricket}. A location support system allows the user to obtain his location with respect to the anchor points but provides the user with the freedom to discover services based on his/her location rather than advertising his position to the system and letting the system provide the services. Therefore, it is important to further investigate the privacy and security issues of localization.  
\subsection{Cost}
Cost is another major challenge to the adoption of indoor localization. Localization systems might require additional infrastructure and anchor nodes which require additional investment. Furthermore, localization on a large scale is challenging and might require dedicated  servers, databases and some proprietary software. This is an added cost and certainly would cause most of the customers/service providers to avoid using localization services. While cost is a major challenge now,  it can be overcome by using the existing infrastructure such as WiFi, cellular networks or a combination of both. 
\subsection{Lack of Standardization}
Currently, there is no standard or governing set of specifications/rules that can serve as a guide for designing localization and proximity techniques. There is no single wireless technology that is widely accepted as the main technology for future localization systems. As evident from our discussion on the proposed systems in the previous sections, a number of different technologies and techniques have been used for the purpose. However, most of the systems are disjoint and there is no ubiquitous {system} that currently exists.  
This poses significant challenges.  
Therefore, we believe that there is a need for proper standardization of localization. Through standardization, we can set the specifications and also narrow down the technologies and techniques that can satisfy the aforementioned evaluation metrics. We also believe that future communication technologies such as 5G should also consider the significance of localization. Furthermore, there is a need for creating a universal benchmarking mechanism for evaluating an indoor localization system. 
 \subsection{Negative impact on the used technology}
 Since the goal is to obtain a localization system which relies on the existing infrastructure such as WiFi APs to provide its services, it is important to limit its negative impact on the basic purpose of the used technology i.e. providing connectivity to the users. WiFi and other technologies in their design, as of now, do not consider localization. This means that the use of such technologies for localization will impact other aspects of these technologies \cite{vasisht2016decimeter}. 
 Therefore, we believe that the localization systems should be designed in an optimal way so that the main functionality of the technologies should not be affected. This might require modifying the existing standards to consider localization so that they can provide indoor localization-as-a-service (ILPaaS). 
\subsection{Handovers}
Due to the wide scale use of different technologies such as WiFi, cellular, Bluetooth, UWB, RFID etc., we believe that the future networks will be highly heterogeneous. Therefore, it is highly likely that the localization system will be a hybrid system that might rely on a number of technologies. To obtain improved performance, there might be a need for \textit{vertical handover} among the RNs which use different technologies. This can be because a certain RN might result in the LOS that improves accuracy. Even if the system relies on a single technology, the limited range of the technology might necessitate \textit{horizontal handover} among different RNs as, in the absence of handovers, the system will not work if the RNs and the user device are out of each others range. While handovers have been extensively studied, the stringent latency and limited resources of localization pose additional challenges. The handover should be done quickly to allow the system to perform efficiently without the user facing any problems. Novel handover algorithms and procedures are required that are less complex (so as to reduce the energy consumption) and able to satisfy the system demands. 
Table \ref{tab:challenges} summarizes the aforementioned challenges to localization  along with the proposed solutions. 

\par As evident from above discussion, localization is going to play an important role in {the} future particularly after the advent of IoT and wide-scale use of communication devices. However, for that to happen, there is a need to optimize existing networks from localization perspective. Different technologies should take into account localization as an important service. For example, realizing the significance of Machine Type Communication and IoT, 3GPP standardization now has dedicated bearers for such communication.  
We believe localization should also be taken into account in such a manner. As an example, if WiFi is to be used for localization, specific mechanisms are needed so that WiFi APs can be used for localization without jeopardizing its primary purpose of connecting different entities.

\section{Conclusions}
\label{sec:conclusions}
In this paper, we have presented a detailed description of different indoor localization techniques (AoA, ToF, RToF, RSSI, CSI etc.) and technologies (WiFi, UWB, Visible Light etc.).The paper also provided a thorough survey of various indoor localization systems that have been proposed in the literature with particular emphasis on some of the recent systems. Using our proposed evaluation framework, the paper evaluated these systems using metrics such as energy efficiency, accuracy, scalability, reception range, cost, latency and availability. We provided a number of use case examples of localization to show their importance particularly after the rise of the IoT and the improved connectivity due to different sensors. The paper also highlighted a number of challenges affiliated with indoor localization and provided general directions and solutions that can help in tackling these challenges.

\section*{Acknowledgment}
This work was in part supported by EPSRC Center for Doctoral Training in High Performance Embedded and Distributed Systems  (HiPEDS, Grant Reference EP/L016796/1), Department of Electrical and Electronics Engineering, Imperial College London, and by the U.S. Army Research Laboratory and the U.K. Ministry of Defence under Agreement Number W911NF-16-3-0001. The views and conclusions contained in this document are those of the authors and should not be interpreted as representing the official policies, either expressed or implied, of the U.S. Army Research Laboratory, the U.S. Government, the U.K. Ministry of Defence or the U.K. Government. The U.S. and U.K. Governments are authorized to reproduce and distribute reprints for Government purposes notwithstanding any copy-right notation hereon. The authors would also like to thank the reviewers for their valuable feedback and comments. 
%

\ifCLASSOPTIONcaptionsoff
  \newpage
\fi

\bibliographystyle{ieeetr}
\bibliography{IEEEabrv,references}

\end{document}